\documentclass[journal=jacsat,
manuscript=article,
layout=onecolumn
]{achemso}

\usepackage{mathrsfs,amsmath,amssymb,bm}
\usepackage{lipsum}
\usepackage{amsfonts}
\usepackage{graphicx}
\usepackage{epstopdf}
\usepackage{makecell, rotating}
\usepackage{multirow}
\usepackage{algorithm, algorithmic}
\usepackage{color}
\usepackage{epsfig,subfigure,dcolumn}
\usepackage{xr-hyper}
\usepackage{float}
\usepackage{enumitem}
\usepackage{hyperref}

\usepackage[capitalize]{cleveref}
\usepackage[section]{placeins}
\usepackage[draft]{changes}
\usepackage{booktabs}
\definechangesauthor[name={Kai Jiang}, color=red]{jk}

\usepackage{xr}
\makeatletter

\newcommand*{\addFileDependency}[1]{
\typeout{(#1)}
\@addtofilelist{#1}
\IfFileExists{#1}{}{\typeout{No file #1.}}
}\makeatother

\newcommand{\br}{\bm{r}}

\newcommand{\bu}{\bm{u}}

\newcommand{\bI}{\boldsymbol{I}}

\newcommand{\calS}{\mathcal{S}}
\newcommand{\calM}{\mathcal{M}}

\allowdisplaybreaks[4]

\usepackage {xr}
\author{Xin Wang}
\affiliation{Hunan Key Laboratory for Computation and Simulation in Science and Engineering, Key Laboratory of Intelligent Computing and Information Processing of Ministry of Education, School of Mathematics and Computational Science, Xiangtan University, Xiangtan, Hunan, China, 411105}

\author{An-Chang Shi}
\email{shi@mcmaster.ca}
\affiliation{Department of Physics and Astronomy, McMaster University, Hamilton, Ontario L8S4M1, Canada}

\author{Pingwen Zhang}
\email{pzhang@pku.edu.cn}
\affiliation{School of Mathematics and Statistics, Wuhan University, Wuhan 430072, China}
\affiliation{School of Mathematical Sciences, Peking University, Beijing 100871, China}

\author{Kai Jiang}
\email{kaijiang@xtu.edu.cn}
\affiliation{Hunan Key Laboratory for Computation and Simulation in Science and Engineering, Key Laboratory of Intelligent Computing and Information Processing of Ministry of Education, School of Mathematics and Computational Science, Xiangtan University, Xiangtan, Hunan, China, 411105}

\title{Stability of diverse dodecagonal quasicrystals in T-shaped liquid crystalline molecules}

\begin{document}
\begin{figure}[H]
	\centering
	\includegraphics[width=14cm]{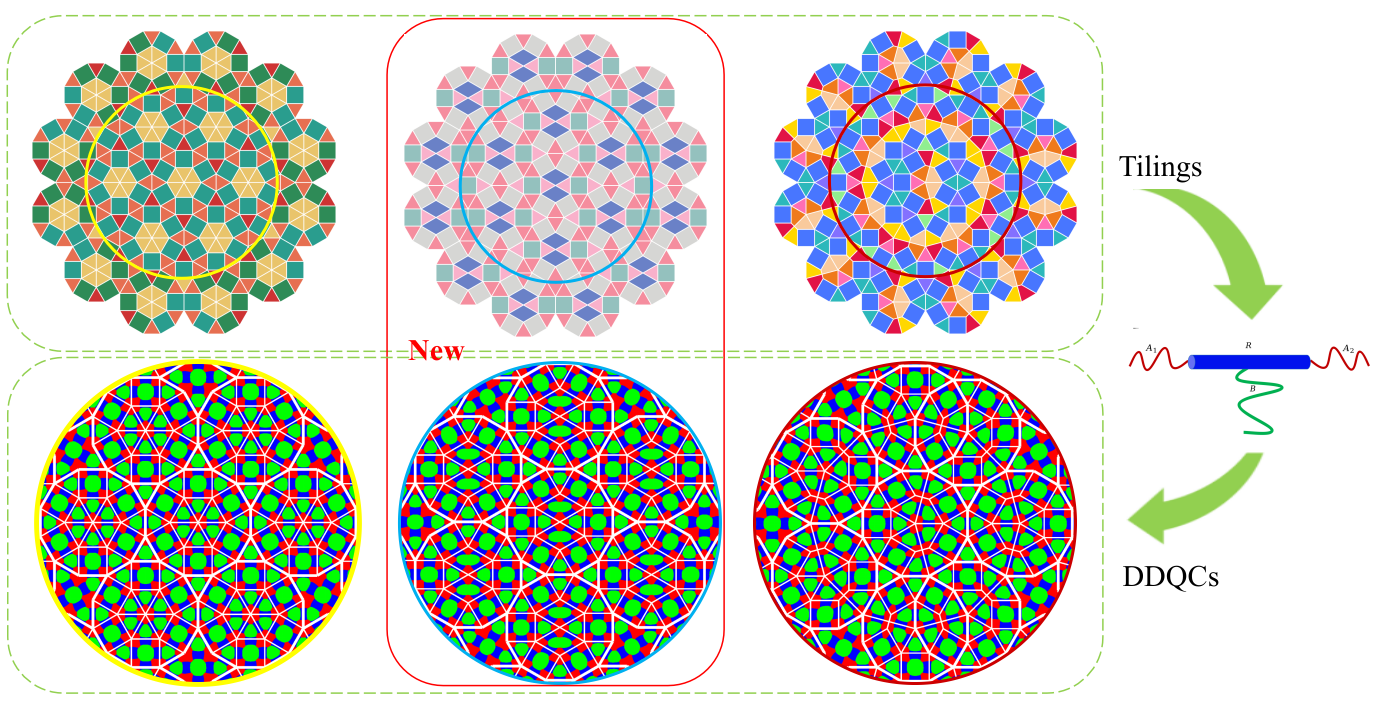}
	\caption*{Table of Contents use only}
\end{figure}

\begin{abstract}
Quasicrystals are intriguing ordered structures characterized by the lack of translational symmetry and the existence of rotational symmetry. The tiling of different geometric units such as triangles and squares in two-dimensional space can result in a great variety of quasicrystals that could be realized by the self-assembly of liquid crystalline molecules. In this study, we introduce three self-similar dodecagonal tilings, including a novel Diamond-Square-Triangle pattern, composed of triangular and quadrangular tiles and examine their thermodynamic stability by using the self-consistent field theory applied to T-shaped liquid crystalline molecules.
Specifically, we detail the inflation rules for the construction of these dodecagonal tilings and analyze their self-similarity, and show that these tilings can be viewed as projections of higher-dimensional periodic lattice points with projection windows. 
Using these dodecagonal tilings as initial configurations of the SCFT results in solutions corresponding to quasicrystals that could form from the T-shaped liquid crystalline molecules. The relative stability of these aperiodic phases is analyzed to obtain design rules that could stabilize quasicrystals. Meanwhile, we provide two criteria for distinguishing three dodecagonal quasicrystals and their approximants by analyzing their diffraction peaks.
These findings shed new lighten on the discovery of new quasicrystals in soft materials.
\end{abstract}

\maketitle

\section{Introduction}
\label{sec:intrd}

Quasicrystals (QCs) are fascinating aperiodic structures that possess rotational symmetry and lack translational symmetry. The discovery of QCs in Al-Mn alloys in 1984 was a breakthrough that changed the perceptions of crystalline order\,\cite{shechtman1984metallic}. Since then, thousands of metallic alloys have been found to exhibit quasicrystalline order\,\cite{steurer2004twenty}. 
Furthermore,  QCs have been observed in a wide range of soft materials, including block copolymers\,\cite{gillard2016dodecagonal,duan2018stability,suzuki2022largest}, liquid crystalline molecules\,\cite{zeng2004supramolecular,zhang2017direct,zeng2023columnar}, nanoparticles\,\cite{talapin2009quasicrystalline}, colloidal particles\,\cite{fischer2011colloidal}, mesoporous
silica\,\cite{terasaki2004mesoporous}, silicon bilayers\,\cite{johnston2011single}, and DNA motifs\,\cite{reinhardt2016self,noya2021design}.
It is interesting to note that the majority of the soft QCs adopt 12-fold rotational symmetry, while only a few of them are with 10-, and 18-fold rotational symmetries\,\cite{nagaoka2018single, liu2022expanding, fischer2011colloidal}.
These soft QCs are quasiperiodic in a plane while homogenous or periodic normal to the plane. Therefore they are regarded as two-dimensional (2D) QCs.
The patterns of 2D dodecagonal quasicrystals (DDQCs) observed experimentally are mainly Square-Triangle (ST) \,\cite{iacovella2011self,  suzuki2022largest} and Quadrangle-Square-Triangle (QST) tilings\,\cite{zeng2023columnar,cao2024under}. Only the ST tiling has been considered theoretically\,\cite{duan2018stability}.
Despite great progresses made over the years, the discovery and stability analysis of diverse DDQCs remain relatively unexplored. 

In general, the study of thermodynamic stability of ordered phases requires a specific physiochemical system that exhibits the desirable phase behaviour and a theoretical framework that allows accurate calculation of the free energy of different phases. Therefore, systems capable of self-assembling into polygonal structures are of great interest for studying soft QCs. For the purpose of studying the emergence and relative stability of various DDQCs that could be constructed from polygonal tiles, a suitable soft matter system is the T-shaped liquid crystalline molecules (TLCMs) shown in Fig.\,\ref{fig:Tmodel}. Extensive experimental studies on this class of molecules have demonstrated their self-assembly into rich polygonal structures, such as Triangle, Dual-Pentagon, Diamond, Square, Pentagon, Hexagon, Octagon, Decagon\,\cite{koelbel2001design,cheng2003calamitic,cheng2004liquld,cook2005supramolecular,chen2005liquid,chen2005carbohydrate,prehm2008distinct,crane2008molecular,cheng2011influence,liu2013dissipative,tschierske2012complex}, and even DDQCs\,\cite{zeng2023columnar}. The rich phase behaviours of the TLCMs make them an ideal platform for studying the relative stability of diverse DDQCs in soft materials.

On the theoretical front, several frameworks have been proposed to study the stability of QCs. One widely used class of theories depends on the construction of multi-length-scale free energy functionals, such as phenomenological Landau-type theories\,\cite{lifshitz1997theoretical, savitz2018multiple} and density functional theories\,\cite{archer2019deriving}. These models are suitable for investigating generic features of QCs and their phase transitions\,\cite{jiang2015stability, subramanian2016three, yin2021transition}. However, these theories usually start from certain hyperthesized multi-length-scale correlation potentials, making it difficult to connect the theory with concrete physical systems. Another widely used class of theories is the self-consistent field theory (SCFT), which is a powerful framework for accurately describing the self-assembly behaviour of inhomogeneous soft materials, particularly polymers and liquid crystal polymers\,\cite{fredrickson2002field, fredrickson2006equilibrium, shi2004self, shi2019self}. Over the past decades, the SCFT has been successfully applied to studying the phase behaviours of various flexible and semiflexible polymer systems\,\cite{morse1994semiflexible, gao2011self-assembly, jiang2013influence, liu2018archimedean}.  The success of SCFT makes it a useful framework to study the phase behaviours of complex molecules, such as the TLCMs.

In this work, we study the emergence and relative stability of various two-dimensional QCs in TLCMs by using dodecagonal aperiodic tilings as initial candidate phases of the SCFT. Starting from the aperiodic tiling theory, we analyze the existing dodecagonal ST tiling and QST tiling and then propose a novel dodecagonal Diamond-Square-Triangle (DST) tiling. The respective inflation rules of these structures are presented in details. Meanwhile, a cut-and-project method is used to analyze these tilings and obtain corresponding projection windows. In the second step of the study, we develop a SCFT framework for the TLCMs. Using the constructed dodecagonal tilings as initial structures, we obtain solutions of the SCFT equations corresponding to three DDQCs as well as other candidate phases. The free energy of these phases are then used to determine their relative stability. Finally, we introduce random DDQCs and obtain SCFT solutions of all ideal and random DDQCs in TLCMs, then perform energy analysis to predict possible ways for stabilizing DDQCs. For three DDQCs and their approximants, we provide two criteria for distinguishing them by analyzing their diffraction peaks.

\section{Dodecagonal aperiodic tilings}

The main task of SCFT study is to obtain solutions of the SCFT equations. Almost all the numerical methods of solving the SCFT equations are iterative in nature, thus the solutions depend crucially on the initial configurations. Over the years, various methods have been developed to construct initial configurations, leading to solutions corresponding to various periodically ordered phases\,\cite{xu2013a}.
However, there have been few studies on the construction of initial configurations for QCs. We propose to use aperiodic tilings to construct initial configurations for diverse DDQCs, thereby opening possibilities for their discovery. In what follows, we introduce the ST and QST tilings, and construct a new DST tiling.

\begin{figure*}
	\centering
	\includegraphics[width=0.9\textwidth]{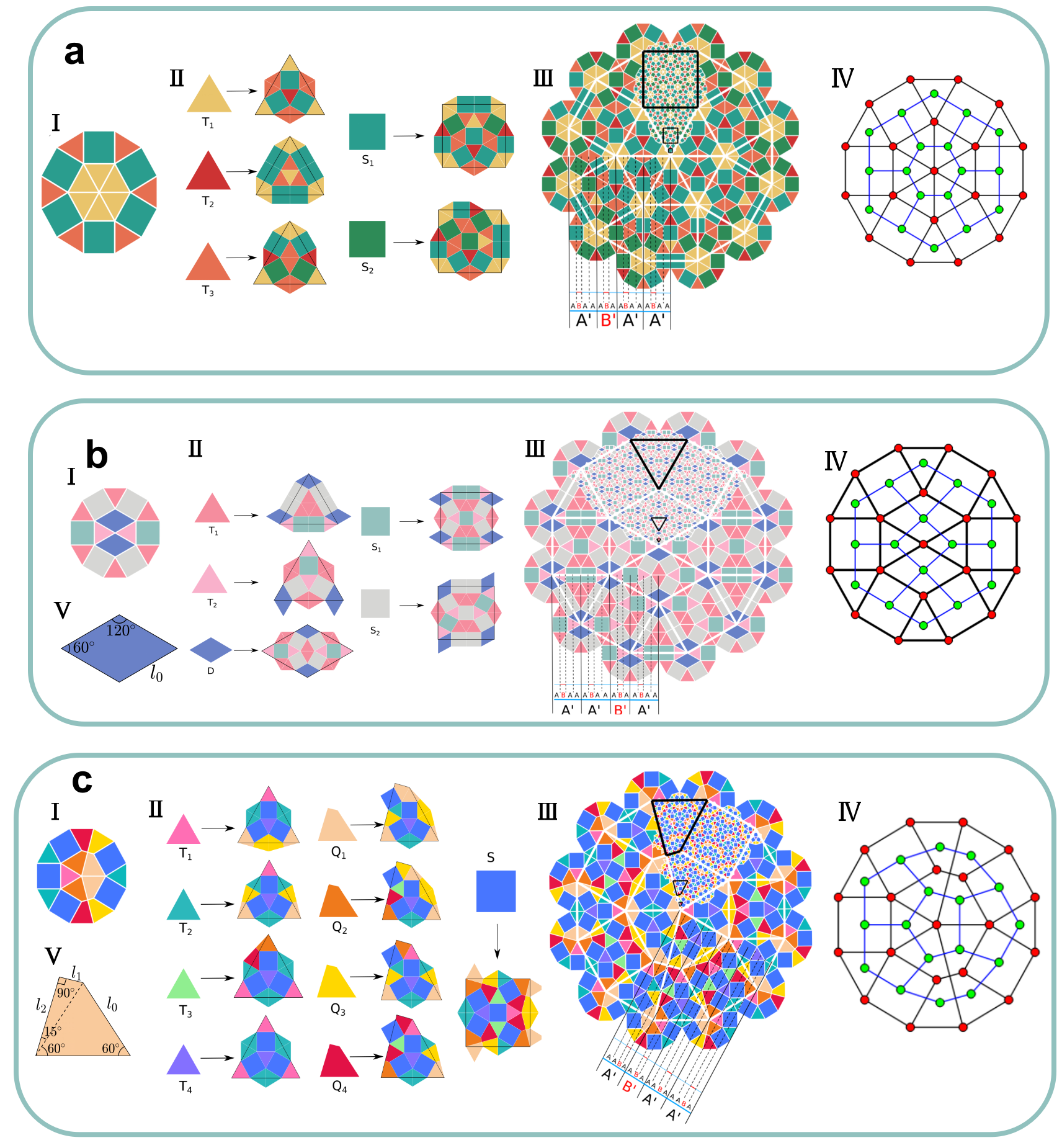}
	\caption{Three dodecagonal aperiodic tilings. (a) Square-Triangle (ST) tiling. (b)
	Diamond-Square-Triangle (DST) tiling. (c) Quadrangle-Square-Triangle (QST) tiling. (I) Fundamental dodecagon. (II) Inflation rules of prototiles. (III) The second-generation tilings of (I). In the	second-generation tilings, parallel line sets show that vertex spacings (``letters") $A$ and $B$ of the given generations construct aperiodic sequences by substituting $\{\rm {A^3B\rightarrow A', A^2B\rightarrow B'}\}$.  The black-outlined parts in the third-generation pattern are self-similar.
 (IV) The schematic tilings of  fundamental dodecagons. The green vertices form  dual tilings of these dodecagonal tilings (red vetices). (V) Diamond and quadrangle tiles.}
\label{fig:ddqcs}
\end{figure*}

The ST tiling is a classical aperiodic tiling consisting of three types of triangles ($T_s$) and two types of squares ($S_s$), all with equal edge lengths. The inflation rule to construct ST tiling is discovered by Stampfli\,\cite{stampfli1986dodecagonal}. 
As shown in Fig.\,\ref{fig:ddqcs}\,a,  the fundamental dodecagon (Fig.\,\ref{fig:ddqcs}\,a(I)) is transformed into the second-generation dodecagon (Fig.\,\ref{fig:ddqcs}\,a(III)) by following the inflation rules shown in Fig.\,\ref{fig:ddqcs}\,a(II). Here the tiles obtained with different inflation rules are distinguished by different colors. The local third-generation pattern in Fig.\,\ref{fig:ddqcs}\,a (III) is obtained by inflating the $T_1$ and $S_1$  twice.
The corresponding edges of the three outlined $S_1$ are magnified by an inflation factor $\alpha = 2+\sqrt{3}$ in each generation, where $\alpha$ is the platinum number, corresponding to the root of $f(x)=x^2-4x+1$.
Meanwhile, the substitution rule $\rm{\{AABA \rightarrow A', ABA \rightarrow B'\}}$ is repeatedly applied to generate an aperiodic sequence.
The ``letters" A and B are two spacings between parallel lines passing through the vertices of the first-generation tiling, and $A'$ and $B'$ are the spacings between parallel lines passing through the vertices of the second-generation tiling.
The length ratios between $A'$ and $A$, $B'$ and $B$ are both $\alpha$. The inflation matrix is given in the Supporting Information (SI), Sec.\,\ref{sec:stif}.
The maximum eigenvalue of the inflation matrix is $\alpha^2$.

The DST tiling, consisting of two kinds of $T_s$, two kinds of $S_s$, and one kind of diamond $D$, is a novel dodecagonal aperiodic tiling constructed in the current study.  All the prototiles have equal edge length, and their inflation rules are given in Fig.\,\ref{fig:ddqcs}\,b (II). 
The shape characteristic of the diamond $D$ is specified in Fig.\,\ref{fig:ddqcs}\,b (V). 
The fundamental dodecagon shown in Fig.\,\ref{fig:ddqcs}\,b (I) containing $6$ $T_1$, $2$ $T_2$, $2$ $D$, $2$ $S_1$ and $4$ $S_2$, can inflate once to form a second-generation tiling, see Fig.\,\ref{fig:ddqcs}\,b (III). Meanwhile, the third-generation pattern is obtained by inflating the first-generation $T_1$, $D$ and $S_2$ twice. The first-, second-, and third-generation $T_s$ are highlighted in black, arranged from bottom to top. Each generation can be magnified by a factor of $\alpha$ relative to the previous generation.
Based on the inflation rule shown in Fig.\,\ref{fig:ddqcs}\,b (II), the number of these tiles in the $(n+1)$-th generation, denoted by $T^{n+1}_1$, $T^{n+1}_2$, $D^{n+1}$, $S^{n+1}_1$ and $S^{n+1}_2$, are
related to those in the $n$-th generation by the inflation matrix $M_{DST}$,
	\begin{equation}
			\label{eqn:MDST}
		\begin{bmatrix}	
		T^{n+1}_1 \\
		T^{n+1}_2\\
		D^{n+1}\\
		S^{n+1}_1\\
		S^{n+1}_2
		\end{bmatrix}
		=
	\begin{bmatrix}
		4 & 3 & 6 & 8 & 8\\
		1 & 2 & 4 & 4 & 4\\
		1 & 1 & 2 & 2 & 2\\
		2 & 1 & 2 & 3 & 2\\
		1 & 2 & 4 & 4 & 5\\
	\end{bmatrix}
	\begin{bmatrix}	
	T^{n}_1 \\
	T^{n}_2\\
	D^{n}\\
	S^{n}_1\\
	S^{n}_2
	\end{bmatrix}
	:=M_{DST}	\begin{bmatrix}	
	T^{n}_1 \\
	T^{n}_2\\
	D^{n}\\
	S^{n}_1\\
	S^{n}_2
	\end{bmatrix}.
	\end{equation}
Here the inflation matrix $M_{DST}$ keeps track of the number of various prototiles in a self-similar way, and its maximum eigenvalue is also $\alpha^2$.
If only the shape ($T$, $D$ and $S$) of prototiles is distinguished, the inflation matrix can be simplified as,
\begin{equation}
	\begin{bmatrix}	
T^{n+1} \\
D^{n+1}\\
S^{n+1}
\end{bmatrix}=
\begin{bmatrix}
	5 & 10 &12 \\
	1 & 2 & 2\\
	3 & 6 &7 \\	
\end{bmatrix}
\begin{bmatrix}
	T^{n} \\
	D^{n}\\
	S^{n}
\end{bmatrix}.
\end{equation}
Furthermore, the vertices of the given generation tilings lying on the set of parallel lines can form two spacings (``letters'') $A$ and $B$, as shown in Fig.\,\ref{fig:ddqcs}\, b (III).
These letters constitute an aperiodic sequence following the substitution rule $\{\rm {ABAA\rightarrow A',
ABA\rightarrow B'}\}$, and the length ratio between $A'$ and $A$, $B'$ and $B$ are both given by $\alpha$.

The prototiles of either ST or DST tilings are equilateral polygons. By introducing an incongruent quadrangle (Fig.\,\ref{fig:ddqcs}\,c (V)), we can construct the QST tiling, where the fundamental dodecagon is illustrated in Fig.\,\ref{fig:ddqcs}\,c (I).  Fig.\,\ref{fig:ddqcs}\,c (II) displays nine prototiles and the corresponding inflation rules, containing four kinds of triangles $T_1$, $T_2$, $T_3$, $T_4$, four kinds of quadrangles $Q_1$, $Q_2$, $Q_3$, $Q_4$ and one kind of square $S$. A second-generation QST tiling with the third-generation of $1$ $Q_2$, $1$ $Q_4$ and $1$ $S$ is shown in Fig.\,\ref{fig:ddqcs}\,c (III). Three black-outlined $Q_4$ are self-similar, which are magnified by $\alpha$ from bottom to top. The ``letters" $A$ and $B$ form the same aperiodic sequence following the substitution rule $\{\rm {ABAA\rightarrow A',
ABA\rightarrow B'}\}$ as DST tiling. 
By tracking the changes in the number of $T_1$, $T_2$, $T_3$, $T_4$, $S$, $Q_1$, $Q_2$, $Q_3$, $Q_4$ based on inflation rules, we can obtain the inflation matrix $M_{QST}$,
\begin{equation*}
\label{eqn:MQST}
	M_{QST}=\begin{bmatrix}	
	1 &  1 & 2   &3 & 1 & 0 & 0 & 1 & 1\\
    2 &  2 & 5/2 &3 & 2 & 2 & 1 &5/2&3/2\\
    0 &  0 & 0   &0 & 2 & 0 & 1 & 0 & 1\\
    1 &  1 & 1   &1 & 1 & 1 & 1 & 1 &1\\
    3 &  3 & 3   &3 & 5 & 7/2 & 7/2& 7/2 &7/2\\
    2m& 2m & 0   &0 & 4n & 2+2m & 2m & 1+m & m\\
    0 &  0 & m   &0 & 4 & 0 & 1 & 1 & 1\\
    m &  m & 0   &0 & 2m+4n & 1+m & 1+m & 1+m/2 & m/2\\
    0 &  0 & m/2 & 0 &4 & 0 & 1 & 0 &  2\\
\end{bmatrix}.
\end{equation*}
During the inflation process, the contribution of $Q$ in all prototiles manifests not only as a whole but also in the form of divided equilateral triangle and right triangle, as indicated by the dashed lines in 
Fig.\,\ref{fig:ddqcs}\,c (V).
Thus the area ratios $m=2\sqrt{3}/(1+2\sqrt{3})$ and $n=1/(1+2\sqrt{3})$ of the two parts relative to $Q$ are used to describe the contribution of $Q$, respectively.
The  maximum eigenvalue of $M_{QST}$ is also $\alpha^2$.

\begin{figure*}[htbp!]
	\centering
	\includegraphics[width=1\textwidth]{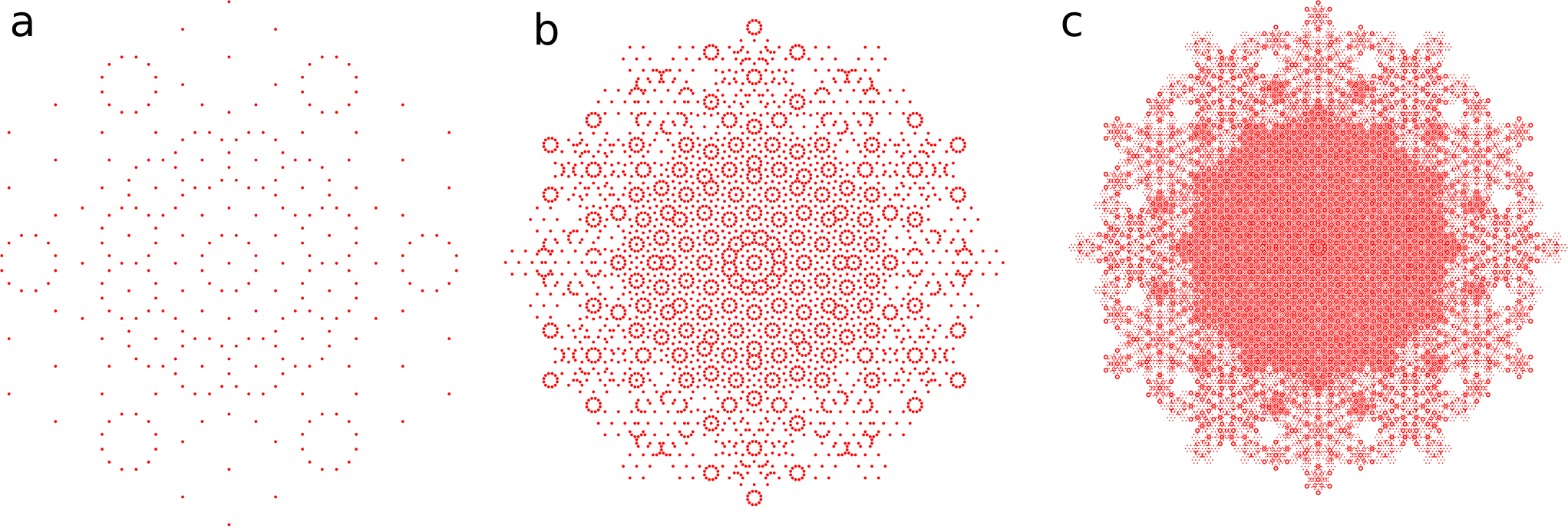}
	\caption{\textbf{
		Higher-dimensional analysis of ST tilings}. a–c, Projection windows obtained
by mapping second- to fourth-generation tilings to the perpendicular
space. The numbers of points are 289, 3727 and 50749, respectively.}
\label{fig:stqcwindow}
\end{figure*}

\begin{figure*}[htbp!]
	\centering
	\includegraphics[width=1\textwidth]{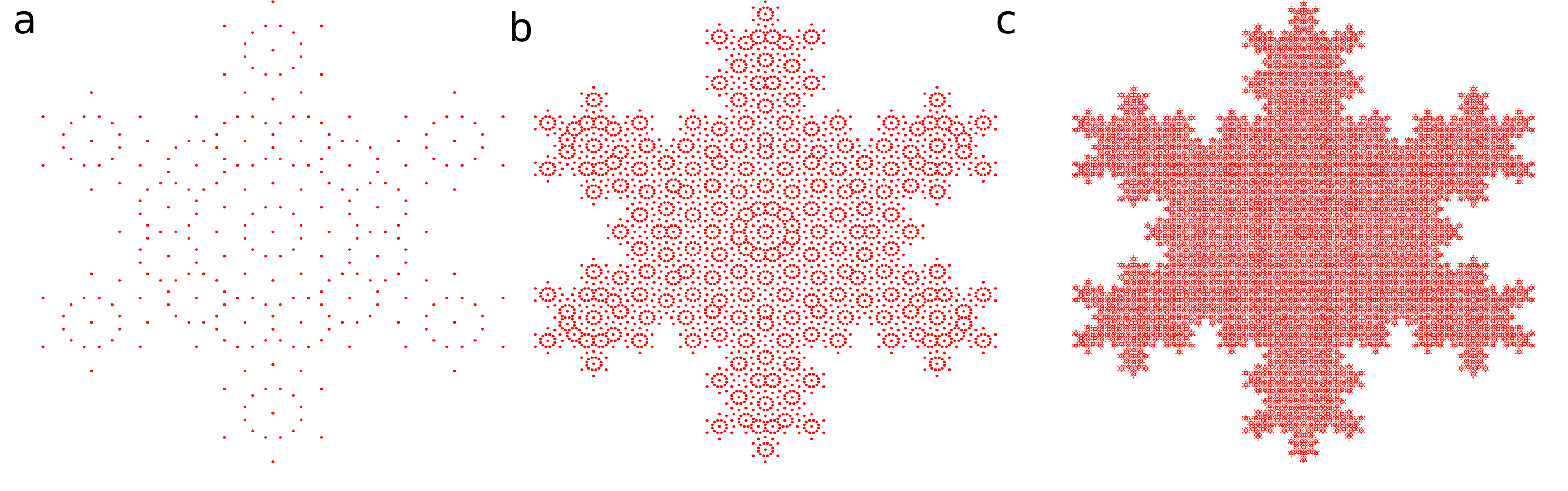}
	\caption{\textbf{
		Higher-dimensional analysis of DST tilings}. a–c, Projection windows obtained
by mapping second- to fourth-generation tilings to the perpendicular
space. The numbers of points are 289, 3767 and 51273, respectively.}
\label{fig:dstwindow}
\end{figure*}

\begin{figure*}[htbp!]
	\centering
	\includegraphics[width=1\textwidth]{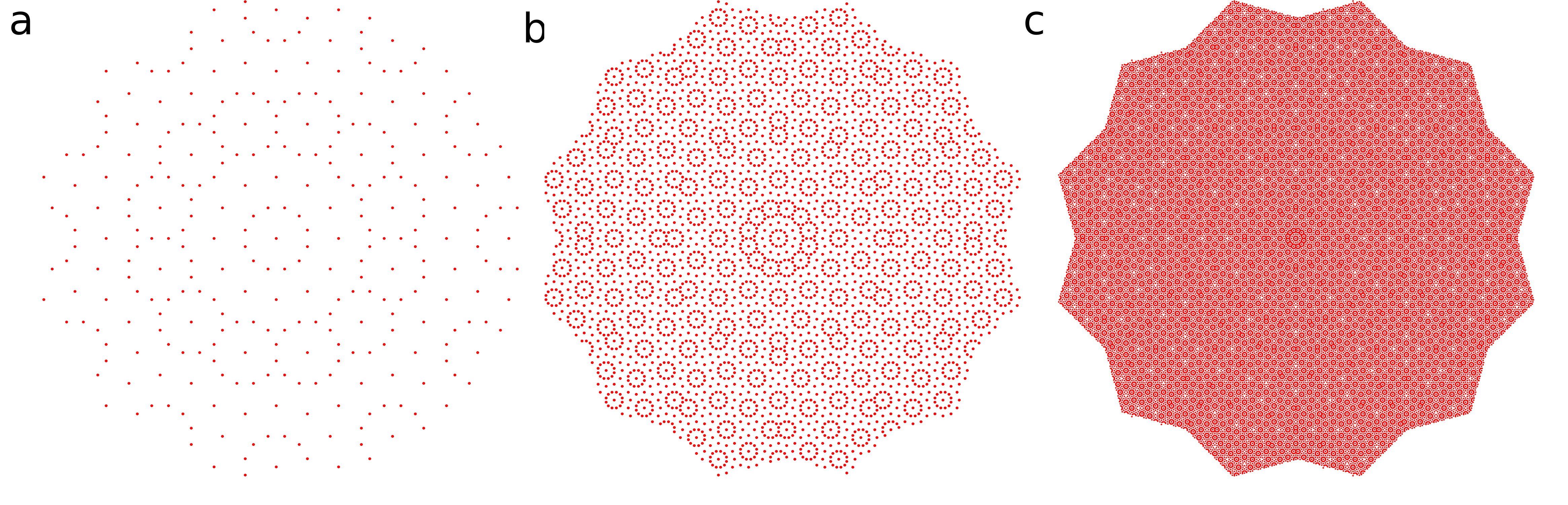}
	\caption{\textbf{
		Higher-dimensional analysis of QST tilings}. a–c, Projection windows obtained
by mapping second- to fourth-generation tilings to the perpendicular
space. The numbers of points are 308, 3956 and 53692, respectively.}
\label{fig:qstwindow}
\end{figure*}

To better understand these dodecagonal tilings, a high-dimensional analysis is carried out by using the classical cut-and-project method\,\cite{meyer1972algebraic}. This involves representing a low-dimensional quasicrystal as a projection of a high-dimensional crystal onto the two-dimensional physical space. The high-dimensional space is divided into parallel and perpendicular spaces. By mapping the vertices of the second-, third-, fourth-generation tilings to the perpendicular space, we construct their projection windows shown in  Fig.\,\ref{fig:stqcwindow},\,\ref{fig:dstwindow},\,\ref{fig:qstwindow}, respectively. These windows exhibit fractal characteristics. Among these windows, the DST tiling window resembles a Koch snowflake and the QST tiling window resembles the shape of two overlapping hexagons. The Hausdorff dimension of the projection window for DST tiling is $\log 6/\log(2+\sqrt{3})\approx1.3605$, while the Hausdorff dimensions of the projection windows for the ST and QST tilings are still unknown.

\section{SCFT of T-shaped liquid crystalline molecules}

For a given molecular system, we solve the SCFT equations using various initial configurations including the dodecagonal tilings to obtain solutions corresponding the different ordered phases. The free energy of these phases is then used to analyze their relative stability. The first step of the SCFT procedure is to develop a SCFT framework for the given molecular systems.

\begin{figure}[htbp] 
	\centering
	\includegraphics[width=6cm]{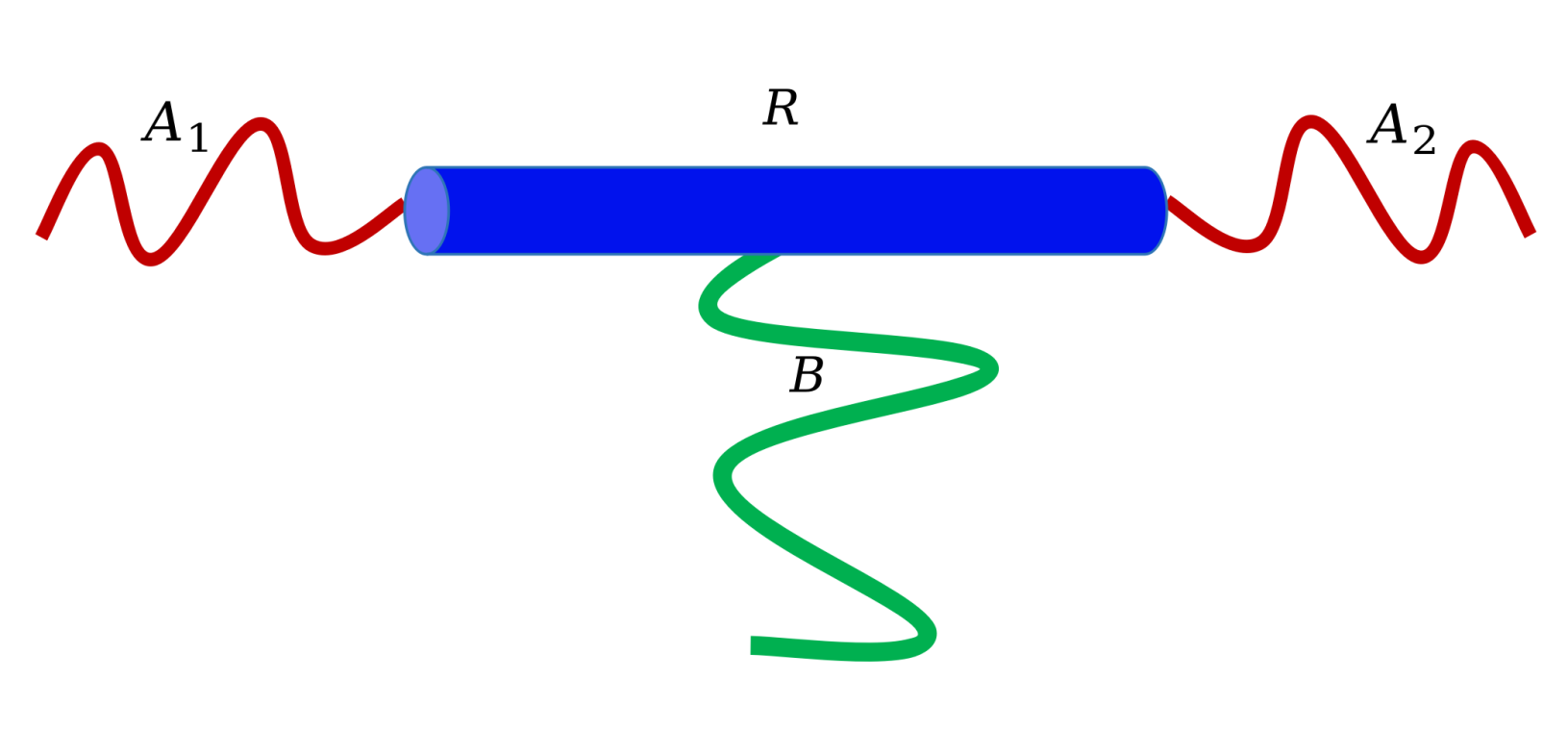} 
	\caption{T-shaped liquid crystalline molecule consisting of a rigid rod $R$ (blue) with two flexible blocks, $A_1$ and $A_2$ (red), tethered at its ends and a flexible side block $B$ (green).} 
	\label{fig:Tmodel} 
\end{figure}

In the current study, we consider an incompressible melt of $n$ T-shaped pentablock terpolymers (Fig.\,\ref{fig:Tmodel}) with an overall degree of polymerization or number of segments $N$ in a volume of $V$.
Each TLCM consists of a liquid crystalline (rigid rod) middle block $R$ (blue) with two flexible end blocks, $A_1$ and $A_2$ (red), tethered at the ends of the rod and one flexible side block $B$ (green) grafted at the middle of the rod. The grafting point of the $B$ block divides the rod $R$ block into two, $R_1$ and $R_2$, subblocks. Effectively the TLCM is a pentablock copolymer with five blocks, $\alpha=\{A_1, A_2, B, R_1, R_2\}$. 
The volume fraction of the $\alpha$-block is denoted by $f_{\alpha}$ with $\alpha=\{A_1, A_2, B, R_1, R_2\}$. For the TLCM specified in Fig.\,\ref{fig:Tmodel} , the volume fractions are given by, $f_{A_2}=\kappa f_{A_1}, f_{R_2}=\gamma f_{R_1}$, with $\sum_{\alpha} f_{\alpha}=1$.

We model the flexible blocks as Gaussian chains and the rigid blocks as wormlike chains. Within the SCFT framework, the free energy per TLCM in unit of thermal energy $k_BT$, where $T$ is the temperature and $k_B$ is the Boltzmann constant, can be expressed as\,\cite{fredrickson2002field},
\begin{equation}
\begin{aligned}
f&=\frac{F}{nk_BT}\\
&=
\frac{1}{V}\int_V
\left[\frac{1}{4N\zeta_{1}}\mu_{1}^2(\br)+\frac{1}{4N\zeta_{2}}\mu_{2}^2(\br)-\mu_{+}(\br)\right]~d\br\\
	&+\frac{1}{2\eta NV}\int_V \calM(\br):\calM(\br)~d\br -\log Q,
\label{H}
\end{aligned}
\end{equation} 
where the Maier-Saupe interaction parameter $\eta$ quantifies the strength of the orientational interaction favoring the alignment of the rod blocks, $Q$ is the single chain partition function. The symbol $A:B$
 denotes the double dot product, defined as $A : B = \sum_{i,j} A_{ij} B_{ij}$.
$\mu_{1}$, $\mu_{2}$ are the general `exchange chemical potential' of the system. $\mu_{+}$ is the pressure potential to ensure the local incompressibility. $\calM$ is the orientational field of semiflexible chain. The parameters $\zeta_1$ and $\zeta_2$ are related to the interactions between the different blocks,
\begin{align*}
\begin{split}
\zeta_{1}=\frac{2\chi_{AB}\chi_{AR}+2\chi_{AB}\chi_{BR}+2\chi_{AR}\chi_{BR}  -\chi_{AB}^2-\chi_{AR}^2-\chi_{BR}^2}{4\chi_{AB}},\quad \zeta_{2}=\chi_{AB},
\end{split}
\end{align*}
where $\chi_{\alpha \beta}$ is the Flory-Huggins interaction parameter between the $\alpha$ and $\beta$ blocks. The mean fields for the different blocks, $w_{\alpha}(\br)$, $\alpha\in\{A,B,R\}$, are linear combinations of the $\mu$-fields,  
$
w_{\alpha}=\mu_{+}-\sigma_{1\alpha}\mu_{1}-\sigma_{2\alpha}\mu_{2},
$
where the $\sigma$-coefficients are given by,
\begin{align*}
\begin{split}
&\sigma_{1A}=\frac{1}{3},\quad\sigma_{1R}=-\frac{2}{3},\quad\sigma_{1B}=\frac{1}{3},\\
&\sigma_{2A}=\frac{1+\alpha}{3},\quad\sigma_{2R}=\frac{1-2\alpha}{3},\quad
\sigma_{2B}=\frac{\alpha-2}{3}, \quad \alpha=\frac{\chi_{AB}+\chi_{AR}-\chi_{BR}}{2\chi_{AB}}.
\end{split}
\end{align*}

The single-chain partition function $Q$, the density distribution of different blocks, $\phi_{\alpha}$ ($\alpha=\{A,B,R\}$), and the orientation order-parameter $\calS$ are computed from the chain propagators,  
$q_{\alpha}(\br, s)$ and $q^{\dagger}_{\alpha}(\br, s)$, $\alpha=\{A_1,A_2,B\}$, $q_{\beta}(\br, \bu,s)$ and $q_{\beta}^{\dagger}(\br, \bu,s)$, $\beta=\{R_1,R_2\}$, which in turn are obtained by solving a set of modified diffusion equations (MDEs). Here $\bu$ is a vector defined on the unit sphere $\mathcal{S}$, which represents the local orientation of the semiflexible molecule. Specifically, we have,
\begin{align}
\begin{split}
Q&=\frac{1}{V}\int_V q_{B}(\br,s)q^{\dagger}_{B}(\br,s)~d\br, \quad \forall s\in[0,f_{B}].\\
 \phi_{A}(\br)&=\frac{1}{Q}\bigg(\int_{0}^{f_{A_{1}}}q_{A_{1}}(\br,s)q^{\dagger}_{A_{1}}(\br,s)~ds +\int_{0}^{f_{A_{2}}}q_{A_{2}}(\br,s)q^{\dagger}_{A_{2}}(\br,s)~ds  \bigg),\\
\phi_{B}(\br)&=\frac{1}{Q}\int_{0}^{f_B} q_{B}(\br,s)q_{B}^\dagger(\br,s)~ds,\\
\phi_{R}(\br)&=\frac{2\pi}{Q}\bigg(\int_{0}^{f_{R_{1}}}\int_{\mathcal{S}} 
q_{R_1}(\br,\bu,s)q_{R_1}^\dagger(\br,\bu,s)~d\bu\,ds+\int_{0}^{f_{R_{2}}}\int_{\mathcal{S}}  q_{R_2}(\br,\bu,s)q_{R_2}^\dagger(\br,\bu,s)~d\bu\,ds\bigg),
\\
\calS(\br)&=\frac{2\pi}{Q}\bigg(\int_{0}^{f_{R_{1}}}\int_{\mathcal{S}} 
q_{R_1}(\br,\bu,s)\big(\bu\bu-\frac{1}{2}\bI\big)
q_{R_1}^\dagger(\br,\bu,s)~d\bu\,ds \\
&+\int_{0}^{f_{R_{2}}}\int_{\mathcal{S}}  q_{R_2}(\br,\bu,s)\big(\bu\bu-\frac{1}{2}\bI\big)q_{R_2}^\dagger(\br,\bu,s)~d\bu\,ds\bigg).
\label{Q}
\end{split}
\end{align}
Here $q_{\alpha}(\br,s)$, $\alpha \in \{A_1,A_2,B\}$ is the forward propagator, representing the probability of finding the $s$-th $\alpha$ segment at a spatial position $\br$ from $s = 0$ to $s = f_{\alpha}$ under mean field $w_\alpha$.
The backward propagator $q_{\alpha}^{\dagger}(\br,s)$ represents the probability weight
from $s = f_{\alpha}$ to $s = 0$. For the Gaussian chains, the propagators satisfy the modified diffusion equations (MDEs)\,\cite{fredrickson2006equilibrium}
\begin{align}
\begin{split}
\frac{\partial}{\partial s}q_{\alpha}(\br,s)=&\nabla_{\br}^2q_{\alpha}(\br,s)-w_{\alpha}(\br)q_{\alpha}(\br,s), \qquad 0\le s\le f_{\alpha},\\
q_{\alpha}(\br,0)=&1.\\
\frac{\partial}{\partial
	s}q_{\alpha}^\dagger(\br,s)=&\nabla_{\br}^2q_{\alpha}^\dagger(\br,s)-w_{\alpha}(\br)q_{\alpha}^\dagger(\br,s),\qquad 0\le s\le f_{\alpha},\\
q_{B}^\dagger(\br,0)=&\int_{\mathcal{S}}  q_{R_1}(\br,\bu,f_{R_{1}})q_{R_2}(\br,\bu,f_{R_{2}})~d\bu.\\
q_{A_1}^\dagger(\br,0)=&\int_{\mathcal{S}}  q_{R_1}^\dagger(\br,\bu,f_{R_{1}})~d\bu,\\
q_{A_2}^\dagger(\br,0)=&\int_{\mathcal{S}}  q_{R_2}^\dagger(\br,\bu,f_{R_{2}})~d\bu.
\end{split}
\label{model:coilPDEplusB}
\end{align}
For semiflexible chains, the forward propagators $q_{\beta}(\br,\bu,s)$, $\beta\in\{R_{1}, R_{2}\}$, 
represent the probability that the endpoint of the $s$ segment at the
spatial position $\br$ and orientational position $\bu$.  They satisfy the ``convective diffusion" equations\,\cite{fredrickson2006equilibrium},
\begin{align}
\begin{split}
\frac{\partial}{\partial
s}q_{\beta}(\br,\bu,s)=&-\nu\bu\cdot\nabla_{\br}q_{\beta}(\br,\bu,s) -\Gamma(\br,\bu)q_{\beta}(\br,\bu,s) + \frac{1}{2\lambda}\nabla_{\bu}^2q_{\beta}(\br,\bu,s), \\
q_{R_1}(\br,\bu,0)=&\frac{q_{A_1}(\br,f_{A_{1}})}{2\pi}, \qquad 0\le s \le f_{R_{1}},\\
q_{R_2}(\br,\bu,0)=&\frac{q_{A_2}(\br,f_{A_{2}})}{2\pi}, \qquad 0\le s\le f_{R_{2}},
\end{split}
\label{model:rodPDER1}
\end{align}
where $\Gamma(\br,\bu)=w_{R}(\br)-\calM(\br):(\bu\bu-\dfrac{1}{2}\bI)$.
The parameter $\nu=(b_{R}/b_{B})(6N)^{1/2}$ measures the size asymmetry of monomer $R$ and $B$, and 
$\lambda$ is the hardness of the semiflexible chain.
The backward propagators of the semiflexible block satisfy,
\begin{align}
\begin{split}
\frac{\partial}{\partial s}q_{\beta}^\dagger(\br,\bu,s)=&\nu\bu\cdot\nabla_{\br}
	q_{\beta}^\dagger(\br,\bu,s)
	-\Gamma(\br,\bu)q_{\beta}^\dagger(\br,\bu,s)
+\frac{1}{2\lambda}\nabla_{\bu}^2q_{\beta}^\dagger(\br,\bu,s),\\
q_{R_1}^{\dagger}(\br,\bu,0)=&\frac{1}{2\pi}q_{B}(\br,f_{B})q_{R_2}(\br,\bu,f_{R_{2}}), \qquad 0\le s\le f_{R_{1}},\\
q_{R_2}^{\dagger}(\br,\bu,0)=&\frac{1}{2\pi}q_{B}(\br,f_{B})q_{R_1}(\br,\bu,f_{R_{1}}),\qquad 0\le s\le f_{R_{2}}.
\end{split}
\label{model:rodPDEplusR1}
\end{align}
Finally, the various fields and densities are related by the self-consistent equations that are obtained by the minimization conditions,
\begin{align}
\begin{split}
&\frac{\delta f}{\delta \mu_+}=\phi_{A}+\phi_{B}+\phi_{R}-1=0,\\
&\frac{\delta f}{\delta \mu_1}=\frac{1}{2N\zeta_{1}}\mu_{1}-\sigma_{1A}\phi_{A}-\sigma_{1R}\phi_{R}-\sigma_{1B}\phi_{B}=0,\\
&\frac{\delta f}{\delta \mu_2}=\frac{1}{2N\zeta_{2}}\mu_{2}-\sigma_{2A}\phi_{A}-\sigma_{2R}\phi_{R}-\sigma_{2B}\phi_{B}=0,\\
&\frac{\delta f }{\delta M} =\frac{1}{\eta N}\calM-\calS=0.\\
\label{Q}
\end{split}
\end{align}

\begin{figure*}
	\centering
	\includegraphics[width=1.0\textwidth]{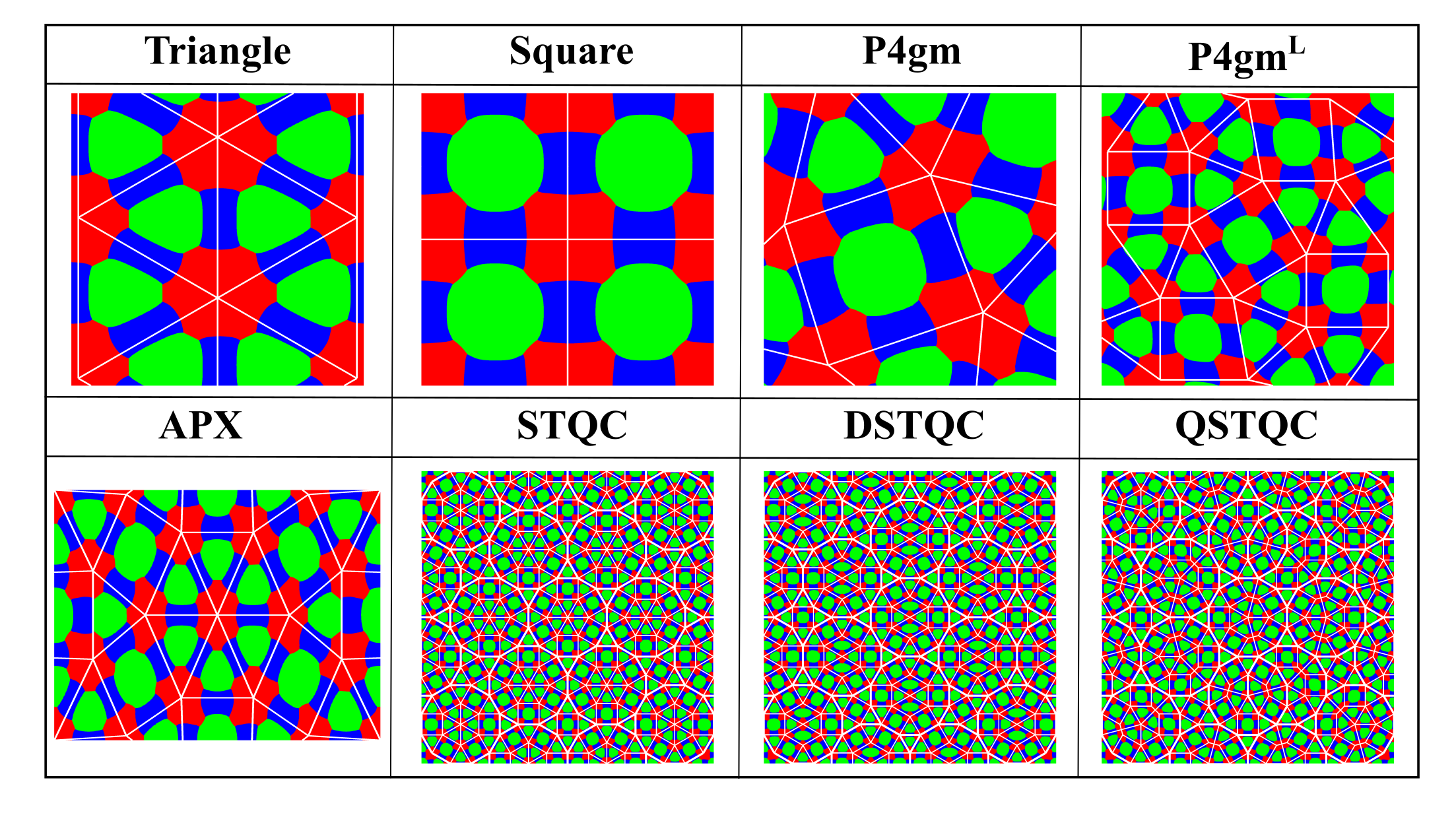}
	\caption{Final configuration density distributions of ordered phases. Red, green and blue colors
		indicate A-, B- and R-block rich domains. The white lines
	connecting neighboring A-domains indicate the corresponding tilings.}
\label{fig:candi}
\end{figure*}

Mathematically, the SCFT of LCMs is a non-local, multi-solution, multi-parameter, high-dimensional, and nonlinear variational problem. The solutions of SCFT are local minima of the free energy landscape corresponding to different ordered structures. Obtaining these solutions numerically requires a diverse set of initial configurations and accurate algorithms. 
In order to study the relative stability of quasicrystals, it is essential to construct initial configurations corresponding to various DDQCs as well as periodic structures. The initial configurations for DDQCs can be obtained by constructing dodecagonal aperiodic tilings. There exists a gap between these aperiodic tilings that is a geometric construction and the density profiles of candidate phases for SCF calculations. To bridge this gap, we employ a method to decorate tilings with smooth functions like Gaussian or $\tanh$ function (SI, Eqn.\,\eqref{eqn:sm}). Specifically, the density functions $\phi_A$ and $\phi_B$ are obtained by decorating the dodecagonal tilings and their dual tilings (red and green vertices in Fig.\,\ref{fig:ddqcs}\, (IV)), respectively, while $\phi_R$ is obtained by using the incompressibility condition. 
Using this approach, the initial configurations of three DDQCs, namely the Square-Triangle DDQC (STQC), Diamond-Square-Triangle DDQC (DSTQC), Quadrangle-Square-Triangle DDQC (QSTQC), with their converged states illustrated in Fig.\,\ref{fig:candi}. 

Accurate and efficient algorithms are required to solve the SCFT equations. In this work, we consider two-dimensional ordered structures and restrict the orientation of the rigid block on a unit circle.
After comparing the accuracy and efficiency of various numerical algorithms, we use the Fourier pseudo-spectral method to discrete spatial and orientational variables, the fourth-order backward difference formula and the fourth-order Runge-Kutta method to discrete the contour variable of Gaussian and Wormlike chain propagators, respectively. And the hybrid nonlinear iteration scheme is used to find the saddle points of the SCFT\,\cite{he2024high}. 
During the self-consistent iteration, we utilize optimization algorithms to relax the calculation domain to make the structures reach their optimal energy states. For more algorithmic details and procedures, please refer to the SI, Sec.\,\ref{sec:smeth}.
Meanwhile, a parallel technique is developed using FFTW-MPI package in C++ language \,\cite{frigo2005design} to accelerate the calculations.

 \section{SCFT Solutions of DDQCs}
 \label{sec:DDQCs}
 
Using the aperiodic tilings as initial configurations allows us to obtain SCFT solutions of three DDQCs shown in Fig.\,\ref{fig:candi}. The self-consistency of all the SCFT solutions are determined by demanding that the free energy difference between consecutive iteration steps is less than the iteration tolerance of $\mbox{tol} =1.0\times 10^{-8}$. The relative stability of all candidate phases is examined by using their free energy.
Phase diagram of the system is constructed by a comparison of the free energies of the candidate phases. The library of candidate phases used in the current study includes the three DDQCs and their crystal approximants, such as P4gm, APX\,\cite{duan2018stability}, $\rm{P4gm^L}$\,\cite{zeng2023columnar}, as well as polygonal structures including Triangle and Square. The corresponding diffraction patterns of each phase are displayed in SI, Figs.\,\ref{fig:qps}-\ref{fig:ps}. 

\begin{figure*}[!hbpt]
	\centering
	\includegraphics[width=1.0\textwidth]{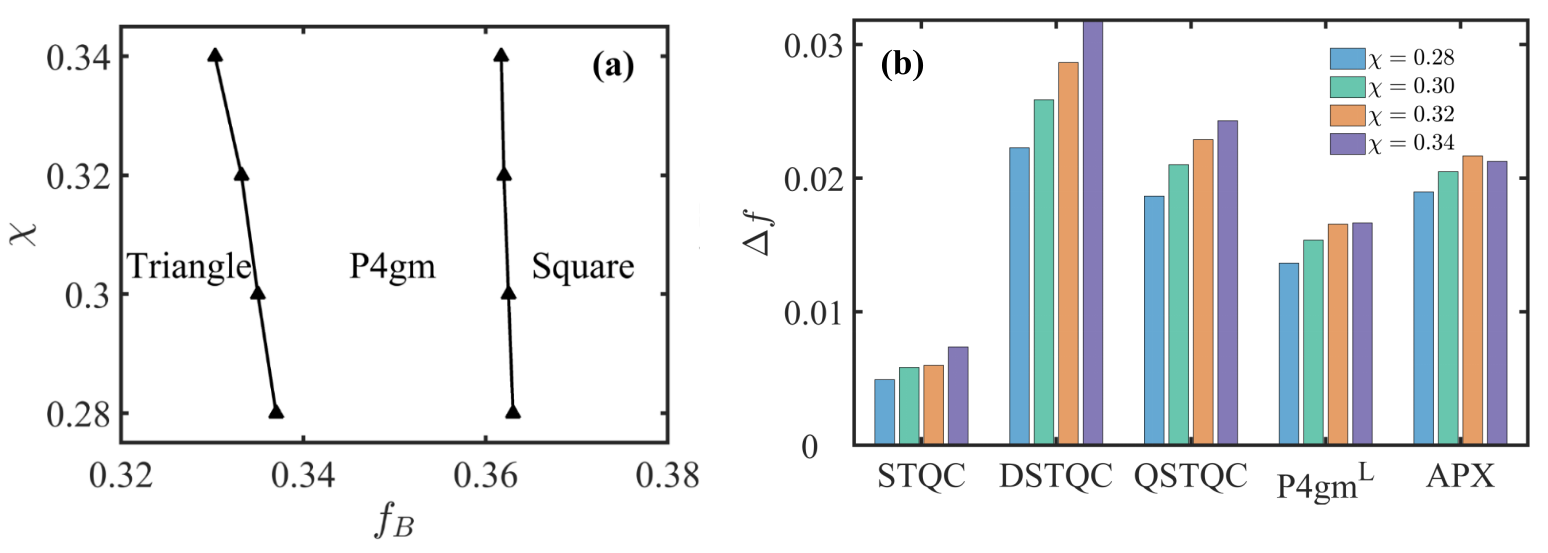}
	\caption{(a) Phase diagram in $\chi-f_B$ plane. Symbols indicate the  transition
	points determined by SCFT, while solid lines are a guide for the eyes. (b)
free energy differences $\Delta f=f-f_{P4gm}$ at
$[\chi-f_B]=[0.28-0.351]$, $[0.30-0.350]$, $[0.32-0.348]$, $[0.34-0.347]$. Here parameters $\kappa=0.4, f_{R_1}=0.14, \gamma=1.1$,  $\eta=0.3$, $\lambda=300$.}
	\label{fig:energy}
\end{figure*}

Previous experiments and simulations have revealed that the DDQCs could form between the stable regions of the triangular and quadrangular phases\,\cite{cook2005supramolecular,chen2005liquid,chen2005carbohydrate,cheng2011influence,liu2013dissipative,tschierske2012complex}. For the case of TLCMs, our previous SCFT calculations\,\cite{he2024theory} have provided a suitable parametric range for the search of stable DDQCs, {\it i.e.} $\kappa=0.4, f_{R_1}=0.14, \gamma=1.1$, $\eta=0.3$, $b_A=b_B=1.0$, and $\lambda=300$. 
Due to the significant impact of Flory-Huggins parameters $\chi_{AR}$, $\chi_{BR}$, $\chi_{AB}$, and the side chain length $f_B$ on phase stability, we employ $\chi=\chi_{AR}$ ($\chi_{BR}= \chi - 0.02$, $\chi_{AB}= \chi-0.04$) to analyze the stability of three DDQCs.
Comparing the SCFT free energy of all ordered phases with fixed parameter, the one with the lowest free energy is taken as the stable structure at that point. The resulting phase diagram in the $\chi$-$f_B$ plane is presented in Fig.\,\ref{fig:energy}\,(a). This phase diagram shows that in the range of $\chi\in [0.28,0.34]$, the DDQCs are metastable phases that do not appear in the phase diagram. A phase transition sequence from Triangle, to P4gm, then to Square is observed when $f_B$ is increased. These stable phases are either congruent triangle/square tilings or their combination (P4gm). The three QCs, STQC, DSTQC, and QSTQC, are all solutions of the SCFT equations but they are metastable states within the parameter range. It is worth to notice that the free energy differences between the DDQCs and their crystalline approximants (P4gm, $\rm{P4gm^L}$) are quite small, at the $10^{-2}$ level, as shown in Fig.\,\ref{fig:energy}(b). 

\begin{figure*}[htbp!]
	\centering
	\includegraphics[width=14.0cm]{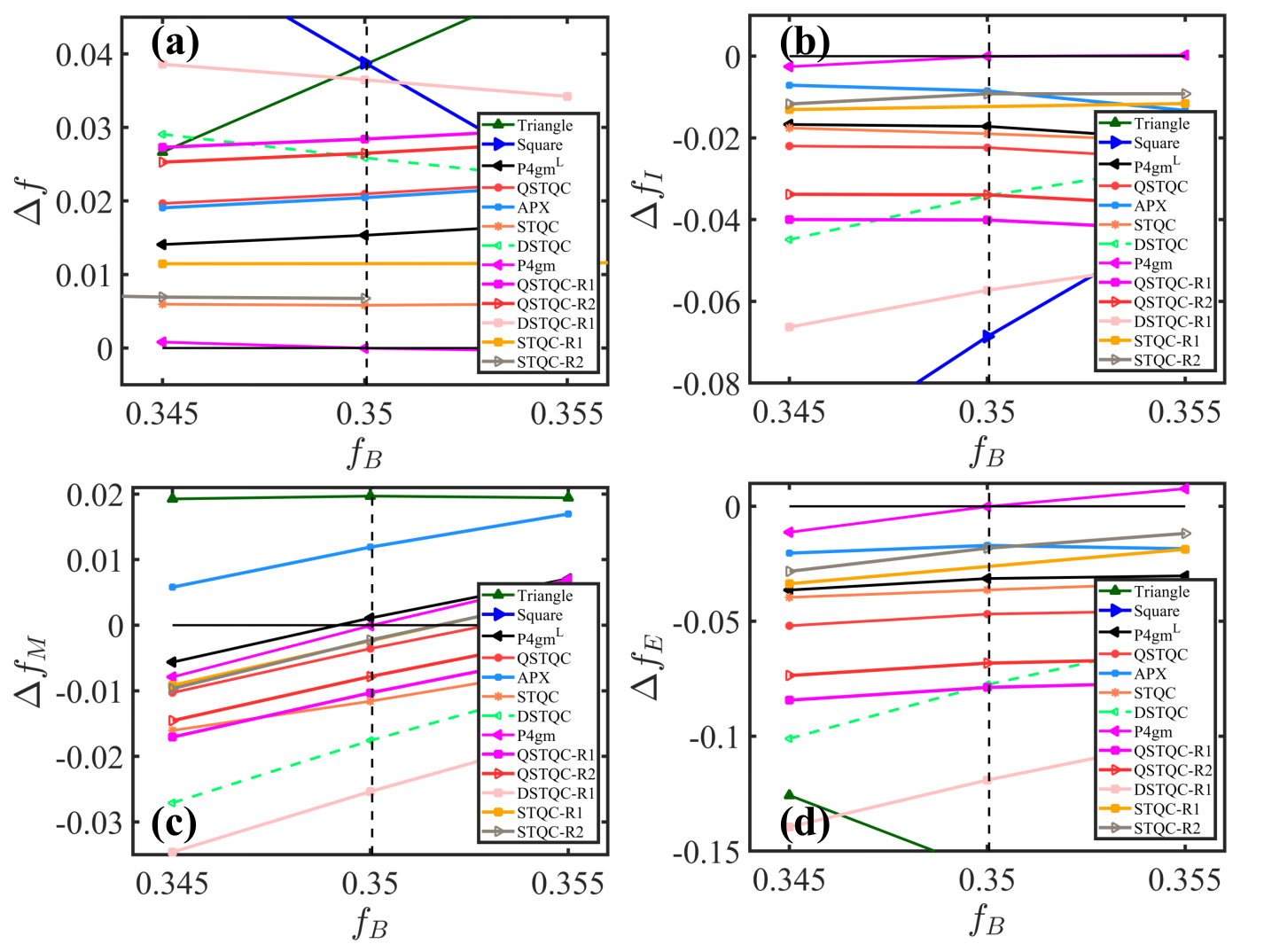}
	\caption{(a)-(d) The free energy difference $\Delta f
=f-f_{P4gm}$ among all phases with $\chi
	=\chi_{AR}=0.30$, where $f_{P4gm}$ is the free energy
of P4gm on the boundary of Triangle and Square indicated by dashed lines. (b)-(d) Three
parts energy differences correspond to (a). Here $\rm{QC-R1/R2}$ representing the random configurations of QCs, are shown in SI, Fig.\,\ref{fig:qps}.}
	\label{fig:en}
\end{figure*}

In order to understand the influence of different factors on the relative stability of the ordered phases, we divide the free energy into three parts: interaction energy ($f_I$), orientation interaction energy ($f_M$) and entropic contribution ($f_E$),
\begin{equation}
\label{eqn:enediv}
\begin{aligned}
f&=
\frac{1}{V}\underbrace{\int_V
	\big(\frac{1}{4N\zeta_{1}}\mu_{1}^2+\frac{1}{4N\zeta_{2}}\mu_{2}^2-\mu_{+}\big)~d\br}_{f_I}\\
&+\underbrace{\frac{1}{2\eta NV}\int_V \calM(\br):\calM(\br)~d\br}_{f_M}
-\underbrace{\log Q}_{f_E}.
\end{aligned}
\end{equation} 
An example of these free energy components with $\chi = 0.30$ and $\eta=0.35$ is shown in Fig.\,\ref{fig:en}. The numerical results from SCFT calculations indicate that the DDQCs have lower $f_I$ and $f_M$ compared to the stable P4gm. However, the relatively larger contribution of $f_E$ prevents the DDQCs become stable phases, despite the advantage brought by the $f_I$ and $f_M$. Thus the entropic contribution $f_E$ dominates the stability of DDQCs. 
Similarly, the DDQCs are metastable phases for the Maier-Saupe parameter in the range of $\eta\in[0.3,0.36]$. The corresponding phase diagram and free energy components (in SI, Sec.\,\ref{sec:chieta}) confirm the dominant role of $f_E$ in stabilizing the different phases.

One interesting question is whether the introduction of randomness to the DDQCs can alter the entropic contribution to the free energy $f_E$, thus enhance their stability. There are two ways to introduce randomness to construct random tilings for the DDQC morphologies.
The first one is the zipper update move approach\,\cite{oxborrow1993random}, which can disrupt the self-similarity through the phase flipping mechanism.
The imperfect aperiodic tilings obtained by this method are almost degenerate compared with the DDQCs, as indicated by the observation that their free energies nearly equal\,\cite{jansen2007aperiodic}. 
This method has already been applied in the construction of random STQC in the SCFT calculations of ABCB tetrablock copolymers, resulting in metastable structures with free energy variations only at the $10^{-3}$ level \,\cite{duan2018stability}. 
The conclusion from these studies is that the zipper update move method could not increase $f_E$ sufficiently to stabilize the DDQCs.
The second one is randomly rotating the fundamental dodecagon in DDQCs such that they still guarantee dodecagonal symmetry, as shown in SI, Fig.\,\ref{fig:qps}. 
This method could either increase or decrease energy $f_E$. For instance, in the STQC-R1 tiling, rotation increases  $f_E$, while in the QSTQC-R1 tiling, it decreases $f_E$.
However, the free energy variations resulting from rotation are also only very small at the $10^{-3}$ level, as shown in Fig.\,\ref{fig:en}. Therefore, the randomness through rotation still does not increase $f_E$ enough to stabilize the DDQCs in the TLCMs.

Based on these SCFT studies, it can be concluded that the stability of the DDQCs obtained via the inflation rules and their randomized variations are insensitive to the parameters $\chi$, $\eta$, $f_B$ of the TLCMs. Therefore, in the future, we should attempt to introduce more blocks or blends into the system to modulate the energy and achieve stable DDQCs.

\section{Distinguish diverse DDQCs and their approximants}
\begin{figure*}[!htbp]
	\centering
	\includegraphics[width=0.8\textwidth]{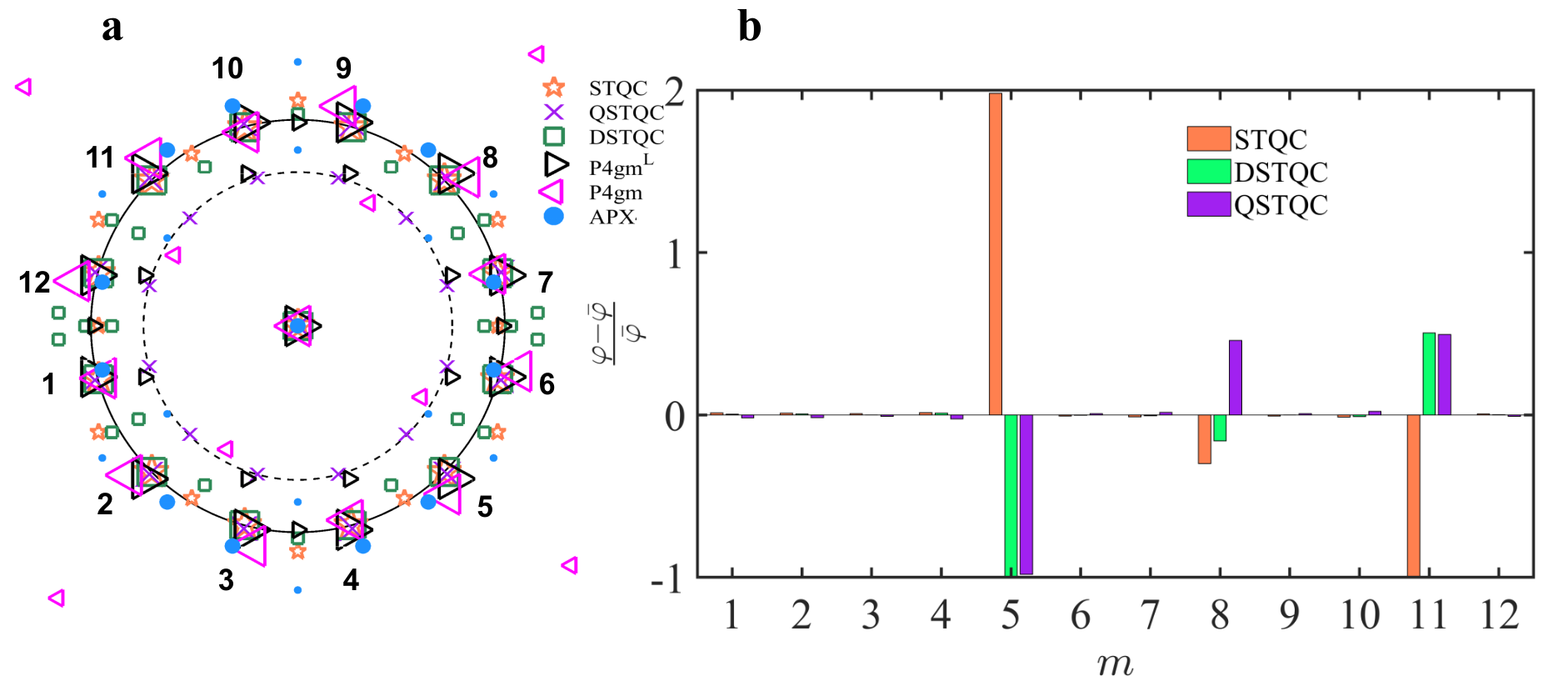}
	\caption{(a) Diffraction peaks of monomer A for DDQCs and their approximants when $\chi=0.30, f_B=0.35$.
	The solid circle represents the location of first-order diffraction
peaks, and the dashed circle shows the location of some second-order diffraction peaks. (b) phase angle differences of three DDQCs. The horizontal axis represents the $m$-th diffraction peak, and the vertical axis represents the phase angle difference between DSTQC, QSTQC and STQC at the first-order diffraction peaks.}
\label{fig:FFT2}
\end{figure*}

The three DDQCs and their approximants exhibit similar symmetries, as shown in Fig.\,\ref{fig:FFT2}\,(a).
The common method of identifying ordered structures through the different symmetries of diffraction peaks \cite{shechtman1984metallic,zeng2004supramolecular} requires further developing. Here, we present two criteria for this task.
We arrange the diffraction peaks in the descending order based on their symmetry and the magnitude of their Fourier coefficients. We designate the top $24$ peaks as the first-order peaks, while the subsequent peaks, ranked from $25$ to $48$, are designated as the second-order peaks.
The first criterion is a comparison of the phase angle at the first-order peaks.
To clearly distinguish the phase angle $\varphi$ of these diffraction peaks, we show only the differences of relative phase angle $\left(\varphi(m)-\bar{\varphi}(m)\right)/\bar{\varphi}(m)$ for three DDQCs when $\chi=0.30, f_B=0.35$ in Fig.\,\ref{fig:FFT2}\,(b) , where $\bar{\varphi}(m)$ represents the average angle of three DDQCs at the $m$-th peak (with the $m$-th largest Fourier coefficient magnitude). STQC can be distinguished from the other two DDQCs at the $m=5, 11$-th peaks, while DSTQC and QSTQC can be distinguished at the $m=8$-th peak.
A full comparison of all the data is provided in SI, Tab.\,\ref{tab:phases}.
The second criterion is a combination of the positions of the first-order and the second-order peaks. 
Focusing on the positions of the A diffraction peaks when $\chi=0.30, f_B=0.35$, as shown in  Fig.\,\ref{fig:FFT2}\,(a), it is evident that the overlapping
first-order peaks (lie on solid circle) are difficult to distinguish between DDQCs and their
approximants, except for APX. However, when combined with non-overlapping second-order peaks positions (lie on dashed circle), it is sufficient to distinguish all structures. More detailed position comparisons are shown in SI, Tab.\,\ref{tab:peaksA}.
The two criteria are presented for the parameter $\chi=0.30,f_B=0.35$ as an example, however, 
they are applicable to other parameters. The reason is attributed to that a variation of parameters only affects the magnitude of the Fourier coefficients of the diffraction peaks.
It is important to note that detecting the phase angle of diffraction peaks might be more challenging in experiments, whereas observing their positions could be easier. Consequently, the second criterion might be more effective for distinguishing DDQCs and their approximants.
\section{Conclusion}
In summary, we constructed three dodecagonal aperiodic tilings, including a novel DST tiling, and used them to construct initial density profiles for the SCFT. By applying the cut-and-project method, we showed that the aperiodic tilings could be obtained from the projection of higher-dimensional periodic lattice points to the lattice points of parallel space within a projection window. The availability of the initial configurations allowed us to obtain accurate solutions of the SCFT equations corresponding to metastable DDQCs.
A free energy analysis underscored the dominant role of the entropic contribution to the free energy in determining the stability of the DDQCs. We also showed that the introduction of random tiling DDQCs is insufficient to stabilize the DDQCs. These insights highlight the need to explore additional mechanisms that can stabilize the DDQCs, such as adjusting molecular topology and introducing extra species via blending.
For three DDQCs and their approximants, we provided two criteria to distinguish them based on diffraction peaks and phase angles. This work represents an effort to investigate DDQCs within the SCFT framework of TLCMs. The methodology and results from the current study not only open new possibilities for research on soft QCs but also provide guidance for experimental investigations aimed at discoverying novel DDQCs.
\begin{suppinfo}
	\begin{itemize}
		\item Aperiodic tilings, Candidate phases, Numerical methods, Effect of $\chi$ and $\eta$ on the thermodynamic stability, Diffraction peaks (PDF)
	\end{itemize}
\end{suppinfo}

\section{Author information}

\textbf{Corresponding author}

\textbf{An-Chang Shi} - \textit{
Department of Physics and Astronomy, McMaster University, Hamilton, Ontario L8S4M1, Canada}

E-mail: shi@mcmaster.ca

\textbf{Pingwen Zhang} - \textit{School of Mathematics and Statistics, Wuhan University, Wuhan 430072, China;
School of Mathematical Sciences, Peking University, Beijing 100871, China}

E-mail: pzhang@pku.edu.cn

\textbf{Kai Jiang} - \textit{ Hunan Key Laboratory for Computation and Simulation in Science and Engineering, Key Laboratory of Intelligent Computing and Information Processing of Ministry of Education, School of Mathematics and Computational Science, Xiangtan University, Xiangtan, Hunan 411105, China}

E-mail: kaijiang@xtu.edu.cn

\noindent\textbf{Notes:} 

The authors declare no competing financial interest.

\bibliography{TQC}
\end{document}


\setlength{\parindent}{2em}
\noindent
This PDF file includes:\\
\ref{sec:AT} Aperiodic tilings \\
\ref{sec:results} Candidate phases\\
\ref{sec:smeth} Numerical methods\\
\ref{sec:chieta} Effect of $\chi$ and $\eta$ on the thermodynamic stability\\
\ref{sec:ddqcs} Diffraction peaks \\
\indent Figures\,\ref{fig:sf}-\,\ref{fig:eeta}\\
\indent Tables\,\ref{tab:N}-\,\ref{tab:peaksA}
\section{Aperiodic tilings} 
\label{sec:AT}
\subsection{The inflation matrix of Square-Triangle tiling} 
\label{sec:stif}

The inflation rule reveals that the numbers of $T_1$, $T_2$, $T_3$, $S_1$ and $S_2$ tiles in the ($n+1$)-th generation denoted by $T^{n+1}_1$, $T^{n+1}_2$, $T^{n+1}_3$, $S^{n+1}_1$ and $S^{n+1}_2$, respectively, 
are related to those in the $n$-th generation by an inflation matrix $M_{ST}$
\begin{equation}
\label{eqn:MST}
\begin{bmatrix}	
T^{n+1}_1 \\
T^{n+1}_2\\
T^{n+1}_3\\
S^{n+1}_1\\
S^{n+1}_2
\end{bmatrix}
=
\begin{bmatrix}
3 & 4 & 3 & 7 & 7\\
1 & 0 & 1 & 2 & 2\\
3 & 3 & 3 & 7 & 7\\
3 & 3 & 1 & 5 & 4\\
0 & 0 & 2 & 2 & 3\\
\end{bmatrix}
\begin{bmatrix}	
T^{n}_1 \\
T^{n}_2\\
T^{n}_3\\
S^{n}_1\\
S^{n}_2
\end{bmatrix}
:=M_{ST}\begin{bmatrix}	
T^{n}_1 \\
T^{n}_2\\
T^{n}_3\\
S^{n}_1\\
S^{n}_2
\end{bmatrix}.
\end{equation}
When the types of $T$ and $S$ are indistinguishable, the relationship between the number of prototiles can be summarized as
\begin{equation}
\begin{bmatrix}	
T^{n+1} \\
S^{n+1}
\end{bmatrix}=
\begin{bmatrix}
7 & 16 \\
3 & 7 \\
\end{bmatrix}
\begin{bmatrix}
T^{n} \\
S^{n}
\end{bmatrix}.
\end{equation}

\subsection{From tilings to initial density function}
\begin{figure}[H]
	\centering
	\includegraphics[width=1\linewidth]{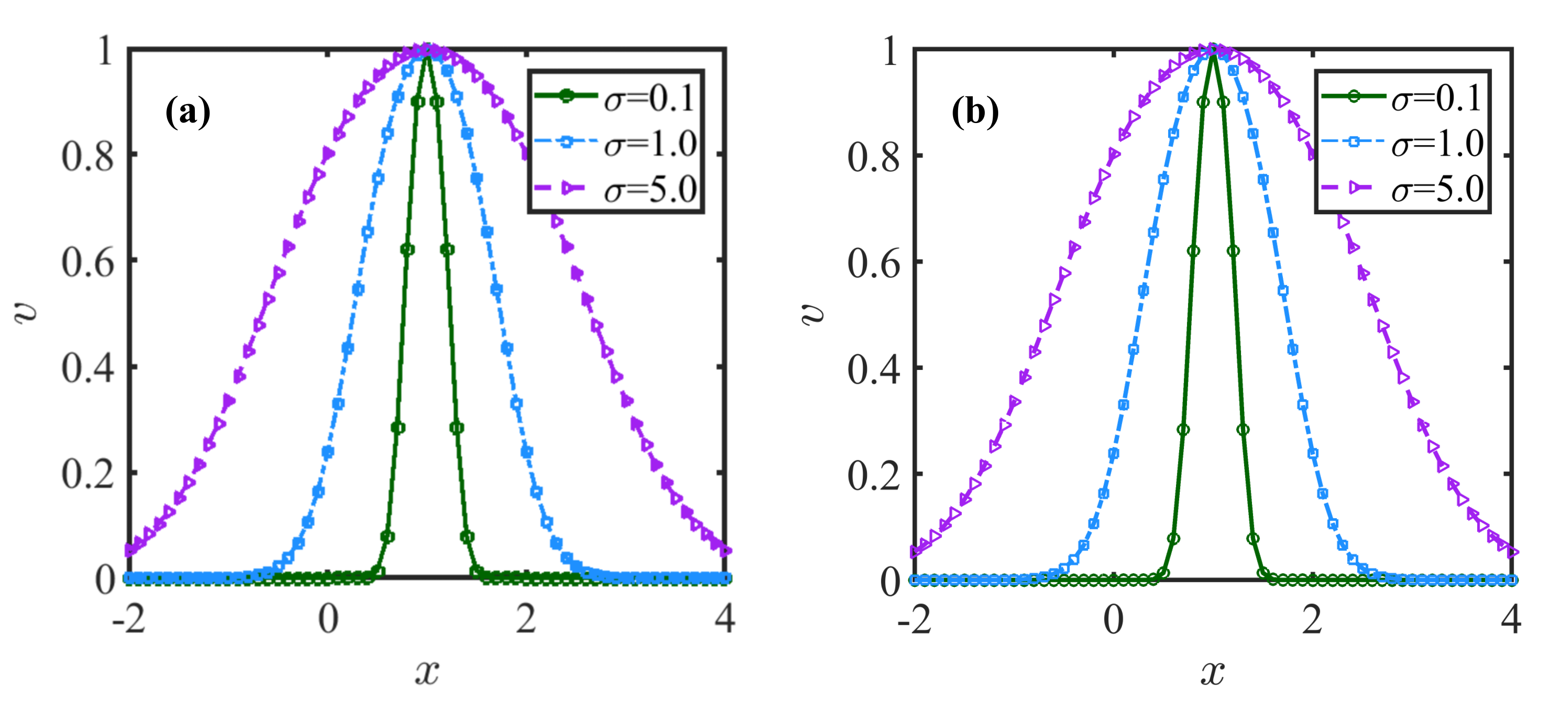}
	\caption{One-dimensional smooth functions  (a).
		$v(x)=\exp\left(-(x-1)^2/\sigma\right)$,
		(b). $v(x)=\left(1-\tanh\left((x-1)/\sigma\right)\right)$ with different $\sigma$.
	}
	\label{fig:sf}
\end{figure}

There exists a gap between the aperiodic tilings that are geometric constructions and the density profiles of candidate phases for SCF calculations. To bridge this gap, we employ a method to decorate tilings with smooth functions like Gaussian or $\tanh$ function. An example is presented here to illustrate the decorating process.
Given a point $(x,y)$ in calculation box $\calA$, the corresponding value $v(x,y)$ can be obtained by all vertices of tiling $(x_i,y_i),i=1,\cdots,N$, $N$ is the number of vertices of tiling. The smooth function $v(x,y)$ can be chosen as
\begin{equation}
\begin{aligned}
\label{eqn:sm}
v(x,y)&=\sum_{i=0}^{N}\exp\left(-\frac{(x-x_i)^2+(y-y_i)^2}{\sigma}\right),\\
\text{or}\\
v(x,y)&=\sum_{i=0}^{N}\left(1-\tanh\left(\frac{(x-x_i)^2+(y-y_i)^2}{\sigma}\right)\right).
\end{aligned}
\end{equation}
The constant $\sigma$ represents the smoothness of the function. This is  illustrated  using  one-dimensional examples in Fig.\,\ref{fig:sf}, where the tiling just corresponds $x=1$.

\section{Candidate phases}
\label{sec:results}
\begin{figure}[H]
	\centering
	\includegraphics[width=1.0\linewidth]{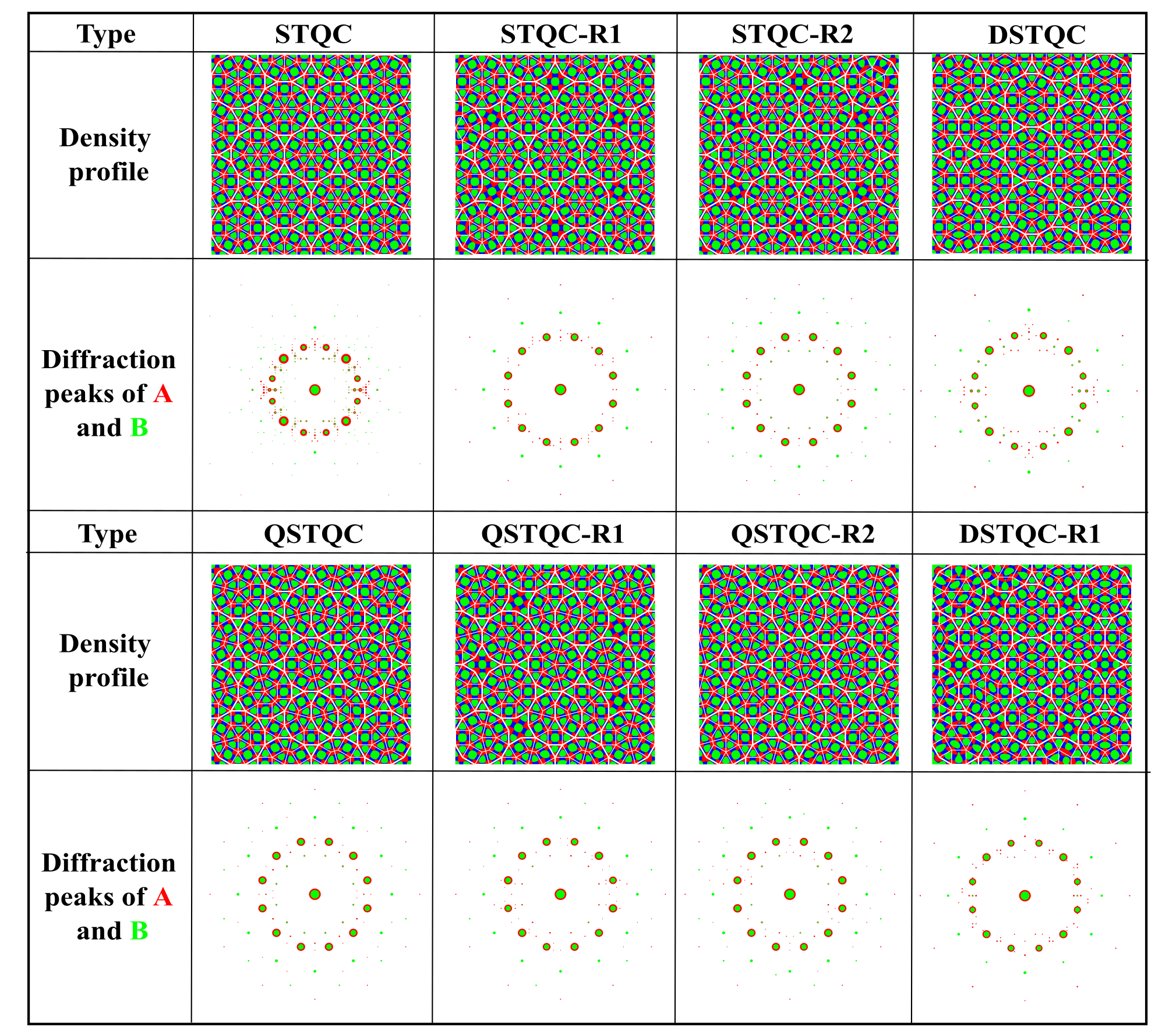}
	\caption{Final configuration density distributions of STQC, DSTQC, QSTQC and their random phases STQC-R1, STQC-R2, DSTQC-R1, DSTQC-R2, QSTQC-R1 (obtained by rotating randomly the first generation dodecagon in the second generation), where A-, B-, R-rich domain are plotted in red, green, blue, respectively. Their diffraction peaks of the subchain A (red) and subchain B (green) all demonstrate the 12-fold symmetry of these DDQCs. The size of green dots is set to be larger than that of red dots to prevent overlap between the two.}
	\label{fig:qps}
\end{figure}

\begin{figure}[H]
	\centering
	\includegraphics[width=1.0\linewidth]{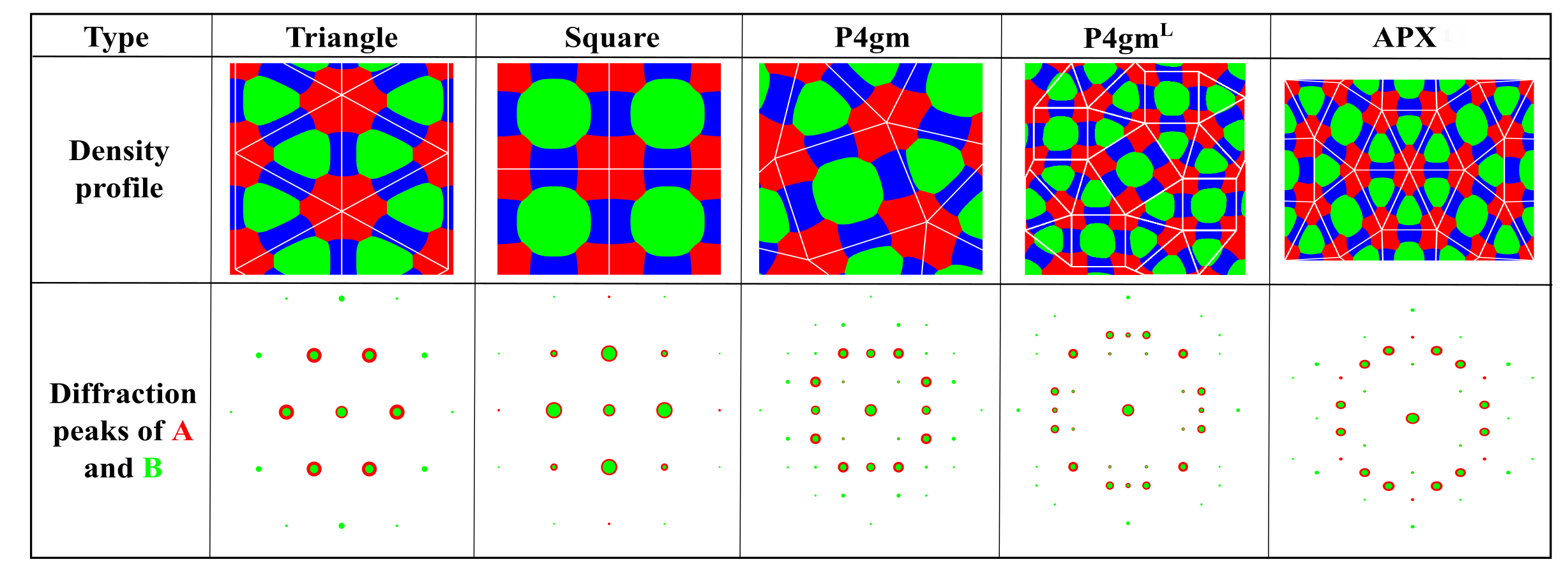}
	\caption{Final configuration density distributions and their diffraction peaks of periodic structures, where A-, B-, R-rich domain are plotted in red, green, blue, respectively. The size of green dots is set to be larger than that of red dots to prevent overlap between the two.}
	\label{fig:ps}
\end{figure}

In this section, we present the final configuration density distributions of all candidate phases. Firstly, we show three DDQCs and their random phases in Fig.\,\ref{fig:qps}, their diffraction peaks of the subchain A (red) and subchain B (green) both demonstrate the 12-fold rotational symmetry of these DDQCs. 
Secondly, we present their  crystal approximants and polygonal phases with diffraction peaks in Fig.\,\ref{fig:ps}. 

\section{Numerical methods}
\label{sec:smeth}
SCFT equations are a set of nonlinear equations with multi-solutions
and multi-parameters. Iterative techniques are required to solve such a complicated system. The specific iteration processes for the self-consistent systems are given in the following:

\begin{itemize}[leftmargin=4em]
   \item[Step 1] Give reasonable initial values of chemical potential fields $\mu_{+}$, $\mu_1$, $\mu_2$,, order-parameter $\bM$, calculation box;
    \item[Step 2] Solve propagators;
    \item[Step 3] Obtain single partition function $Q$, density operators $\phi_{\alpha}, \alpha=\{A,B,R\}$ and order-parameter $\bm S$;
    \item[Step 4] Update potential fields $\mu_{+}$, $\mu_1$, $\mu_2$, order-parameter $\bM$ and calculation box;
    \item[Step 5] If a given convergent condition is achieved, stop iteration procedure, else goto Step 2.
\end{itemize}

In this work, the ternary pentablock T-shaped liquid crystalline molecules are confined on a two-dimensional rectangular box of sizes $L_x \times L_y$ with periodic boundary condition.
The orientational vector $\bu$ is restricted to unit circle. We develop the Fourier pseudo-spectral method and high-order contour scheme to solve SCF equations.
To ensure reliable accuracy, we use enough grids to discretize spatial variable. Concretely, the grid size is set as
$L_x/N_x$ and $L_y/N_y$, the orientation $\bu$ is divided into $N_{\theta}= 16$ points, the chain contour $s$ is divided into $T$ points, as shown in Tab.\,\ref{tab:N}. Meanwhile, we apply the hybrid conjugate gradient method to accelerate the SCF iteration. All numerical methods are presented below\,\cite{he2024high}, respectively. 
\begin{table}[H]
	\centering
	\caption{The number of discrete points and initial calculation domains of all candidate phases, which of the random phases are consistent with the corresponding DDQCs.}
	\begin{tabular}{ccccccccc}
		\toprule
		Phase &  Triangle & Square & P4gm & $\rm{P4gm^L}$ & APX & STQC & DSTQC & QSTQC\\   
		\midrule
		$L_x$ & 7.81 & 7.33 & 6.30 & 13.87 & 16.5 & 47 & 47 & 47 \\
		$L_y$ & 6.53 & 7.33 & 6.30 & 13.46 & 10.8 & 47 & 47 & 47 \\
		$N_x$ & 81   &  81  &  81  & 121   &  143 & 401& 401& 401\\
		$N_y$ & 81   &  81  &  81  & 121   &   81 & 401& 401& 401\\
		$T$     & 200  &  200 &  200 & 300   &  300 & 800& 800& 800\\
		\bottomrule
	\end{tabular}
	\label{tab:N}
\end{table}

\subsection{Fourier pseudo-spectral method}
For spatial function $f(\br)$ with calculation box
$$\calA=(\ba_1, \ba_2)=\begin{bmatrix}
	L_x & 0\\
	0 & L_y\\
\end{bmatrix},
$$
The corresponding primitive reciprocal lattice $\calB =(\bb_1,\bb_2)$, where $\bb_j$ satisfies $\ba_j^T\bb_k =
2\pi \delta_{j,k}, ~j,~k = 1, 2$. 
The rectangular box is divided into a grid lattice of $N_x \times N_y$.

The Fourier series of periodic function $q(\br,\bu,s)$ on discrete
point $\br_{lj}$, orientational position
$u_m=(\cos \theta_m,\sin\theta_m)$, $\theta_m= 2\pi m/ N_\theta,
m=0,\cdots,N_{\theta}-1$, can be written
\begin{align}
	q(\br_{\ell j}, u_m, s)=\sum_{\bk \in \bm{K}, v\in \bm{V}} \hat{q}(\bk,
	v, s) e^{i(\calB\bk)^T \br_{\ell j}} e^{ivu_m}
	\label{Fourier:expansion}
\end{align}
where the discrete Fourier coefficient $\hat{q}(\bk, v,s)$ can be obtained by the Fast
Fourier Transform (FFT).
$ \bm{K}$ and $\bm{V}$ are defined as follows
\begin{equation}
	\begin{aligned}
\bm{K}:=&\{\bk=(k_x, k_y) \in \bbZ^2, ~-N_x/2\leq k_x<N_x/2,~
-N_y/2\leq k_y <N_y/2\},\\
 \bm{V}:=&\{v\in\bbZ:~ -N_{\theta}/2 \leq v <N_{\theta}/2  \} .
\end{aligned}
\end{equation}

We carry out the FFTW-MPI package in C++ language to realize the parallel FFT in the spatial discretization.

\subsection{Fourth-order backward differentiation (BDF4) scheme} 
We apply the BDF4 method to discretize the MDEs of propagators $q_{\alpha}, \alpha=\{A_1, A_2,B\}$ of flexible chains\,\cite{cochran2006stability} 
\begin{align}
\begin{split}
	&\frac{25}{12}q_{\alpha}^{n}(\br)-4q_{\alpha}^{n-1}(\br)+3q_{\alpha}^{n-2}(\br)-\frac{4}{3}q_{\alpha}^{n-3}(\br)+\frac{1}{4}q_{\alpha}^{n-4}(\br)\\
	&=\Delta s\left[\nabla^2
	q^{n}_{\alpha}(\br)-w^n(\br)\left(4q_{\alpha}^{n-1}(\br)-6q_{\alpha}^{n-2}(\br) +4q_{\alpha}^{n-3}(\br)-q_{\alpha}^{n-4}(\br)\right)\right]
\end{split}
\end{align}
The initial values for the first four steps required to apply this formula are obtained using the Richardson’s extrapolation\,\cite{ranjan2008linear}, based on the second-order operator-splitting scheme\,\cite{rasmussen2002improved}. 

\subsection{Fourth-order Runge-Kutta (RK4) method} 
We employ the RK4 method to discretize the semiflexible propagator equations for $q_{\beta}, \beta=\{R_1,R_2\}$, which can be rewritten as 
\begin{align}
\frac{\partial}{\partial s}q_{\beta}(\br,\bu,s)=F(q_{\beta}(\br,\bu,s)),
\label{RK:ODE}
\end{align}
where $F(q_{\beta}(\br,\bu,s))=\nu \bu \cdot \nabla_{\br} q_{\beta}(\br,\bu,s)-
\Gamma(\br,\bu)q_{\beta}(\br,\bu,s)+\dfrac{1}{2\lambda}q_{\beta}(\br,\bu,s)$.
From $s_{n-1}$ to $s_{n}=s_{n-1}+\Delta s$, denote
\begin{equation}
\begin{aligned}
K_{1}&=F(q_{\beta}^{n-1}(\br,\bu)),
\\
K_{2}&=F(q_{\beta}^{n-1}(\br,\bu)+\Delta sK_{1}/2),
\\
K_{3}&=F(q_{\beta}^{n-1}(\br,\bu)+\Delta sK_{2}/2),
\\
K_{4}&=F(q_{\beta}^{n-1}(\br,\bu)+\Delta sK_{3}).
\end{aligned}
\end{equation}
The RK4 discretization scheme can be written as
\begin{align}
q_{\beta}^{n}(\br,\bu)=q_{\beta}^{n-1}(\br,\bu)+\frac{\Delta s}{6}(K_{1}+2K_{2}+2K_{3}+K_{4}).
\end{align}

\subsection{Iteration method} 

The equilibrium ordered structures correspond to saddle-point solutions of SCFT. To find these saddle points, we employ the hybrid conjugate gradient method\,\cite{liang2015efficient, he2024high}. It should be noted that as the conjugate gradient scheme is an algorithm for finding the minimum of a function, we should use it for $\mu_+$, $\mu_-$, $\bM$, $\calB=(b_{ij})_{i,j=1}^{2}$  separately for this saddle-point problem. Specifically, we search for $\mu_+$ along the direction of $\frac{\delta f}{\delta\mu_+}$ and search for $\mu_-$, $\bM$, $\calB$ along the direction of $-\frac{\delta f}{\delta\mu_-}$,  $-\frac{\delta f}{\delta \bM}$, $-\frac{\partial f[b_{ij}]}{\partial b_{ij}}$ (computed by central differential method), respectively. For the minimization direction, 
\cref{alg:hcg} gives the implementation of the hybrid conjugate gradient method. In the algorithm, $\alpha$ is a hybrid factor to adjust the conjugate gradient direction. If $\alpha =0$, the hybrid scheme degenerates to alternative direction iteration scheme, and if $\alpha = 1$, it is the conjugate gradient scheme. By choosing $0 < \alpha < 1$, the hybrid method combines the robustness of the alternative direction iteration method with the efficiency of the conjugate gradient method. 
\begin{algorithm}[H]
	\caption{Hybrid conjugate gradient method}
	\label{alg:hcg}
	\begin{algorithmic}
	\STATE give initial values $\mu_0$, calculate $g_0=\delta f[\mu_0]/\delta \mu_0$, $d_0=-g_0$,\\
	\FOR{$k=0,1,2,\cdots$}
		\STATE choose $\lambda_k$ to minimize $f[\mu_k+\lambda_k d_k]$,\\
		\STATE $\mu_{k+1}=\mu_k+\lambda_k d_k$,\\
		\STATE $g_{k+1}=\delta f[\mu_{k+1}]/\delta \mu_{k+1}$,\\
		\STATE $\gamma_{k+1}=(g^T_{k+1}g_{k+1})/(g^T_kg_k)$,\\
		\STATE $d_{k+1}=-g_{k+1}+\alpha\gamma_{k+1}d_{k}$.
	\ENDFOR
	\end{algorithmic}
\end{algorithm}

\section{Effect of $\chi$ and $\eta$ on the thermodynamic stability}
\label{sec:chieta}
\begin{figure}[htbp!]
	\centering
	\includegraphics[width=12cm]{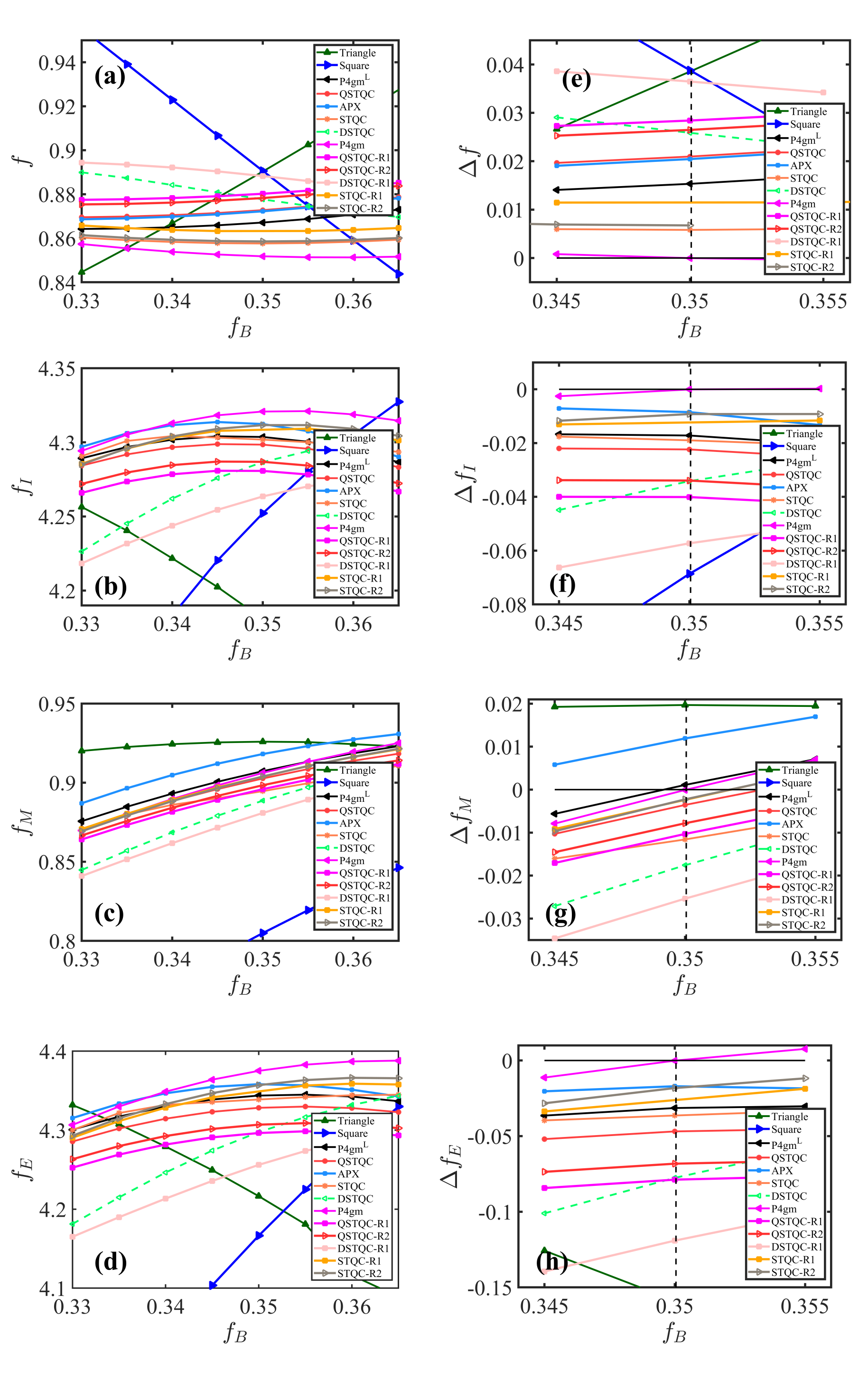}
	\caption{(a)-(d) The free energy comparison among all phases with $\chi
	=\chi_{AR}=0.30$. (e) Corresponding energy difference $\Delta f
=f-f_{P4gm}$ of (a), where $f_{P4gm}$ is the free energy
of Triangle and Square boundary indicated by dashed lines. (f)-(h) Three
parts energy differences correspond (e). }
	\label{fig:en}
\end{figure}

To analyze the dominant factors of stability, we divide the free energy
into three parts: interfacial  energy ($f_I$), orientation
interaction energy ($f_M$) and entropic contribution ($f_E$).
\begin{equation}
\label{eqn:enediv}
\begin{aligned}
f&=
\frac{1}{V}\underbrace{\int_V
	\big(\frac{1}{4N\zeta_{1}}\mu_{1}^2+\frac{1}{4N\zeta_{2}}\mu_{2}^2-\mu_{+}\big)~d\br}_{f_I}\\
&+\underbrace{\frac{1}{2\eta NV}\int_V \calM(\br):\calM(\br)~d\br}_{f_M}
-\underbrace{\log Q}_{f_E}.
\end{aligned}
\end{equation} 
Three contributions of all phases when $\chi = 0.30, \eta=0.35$ are shown in Fig.\,\ref{fig:en} together with the total free energy difference.
For clarity, the energy differences of these quantities at the Triangle/Square phase boundary (equal free energy) are also given.

\begin{figure}[htbp!]
	\centering
	\includegraphics[width=10cm]{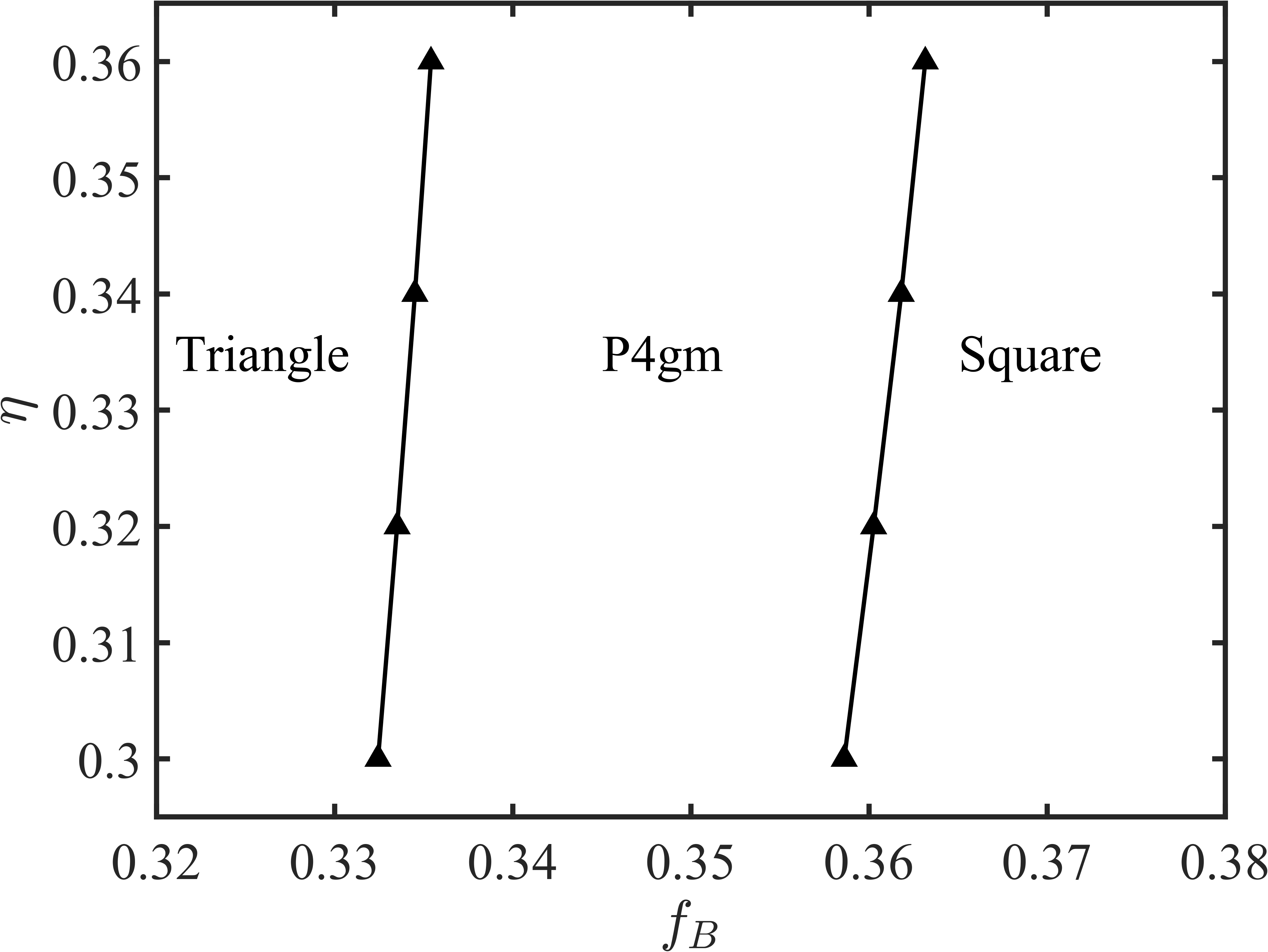}
	\caption{Phase diagram in $\eta-f_B$ plane. Symbols indicate the transition points determined
	by SCFT, while solid lines are a guide for the eyes.}
	\label{fig:etaf}
\end{figure}

The effect of $\eta \in [0.30, 0.36]$ on the thermodynamic stability of DDQCs is also analyzed.
With $\chi = 0.30$, the $\eta-f$ phase diagram is shown in Fig.\,\ref{fig:etaf}, where the phase sequence still remains Triangle $\mapsto$ P4gm  $\mapsto$ Square. This indicates that the perturbation of $\eta$ does not affect the stability of the three DDQCs. Meanwhile, the influence of side chain length on phase stability remains unchanged. 

\begin{figure}[htbp!]
	\centering
	\includegraphics[width=13.0cm]{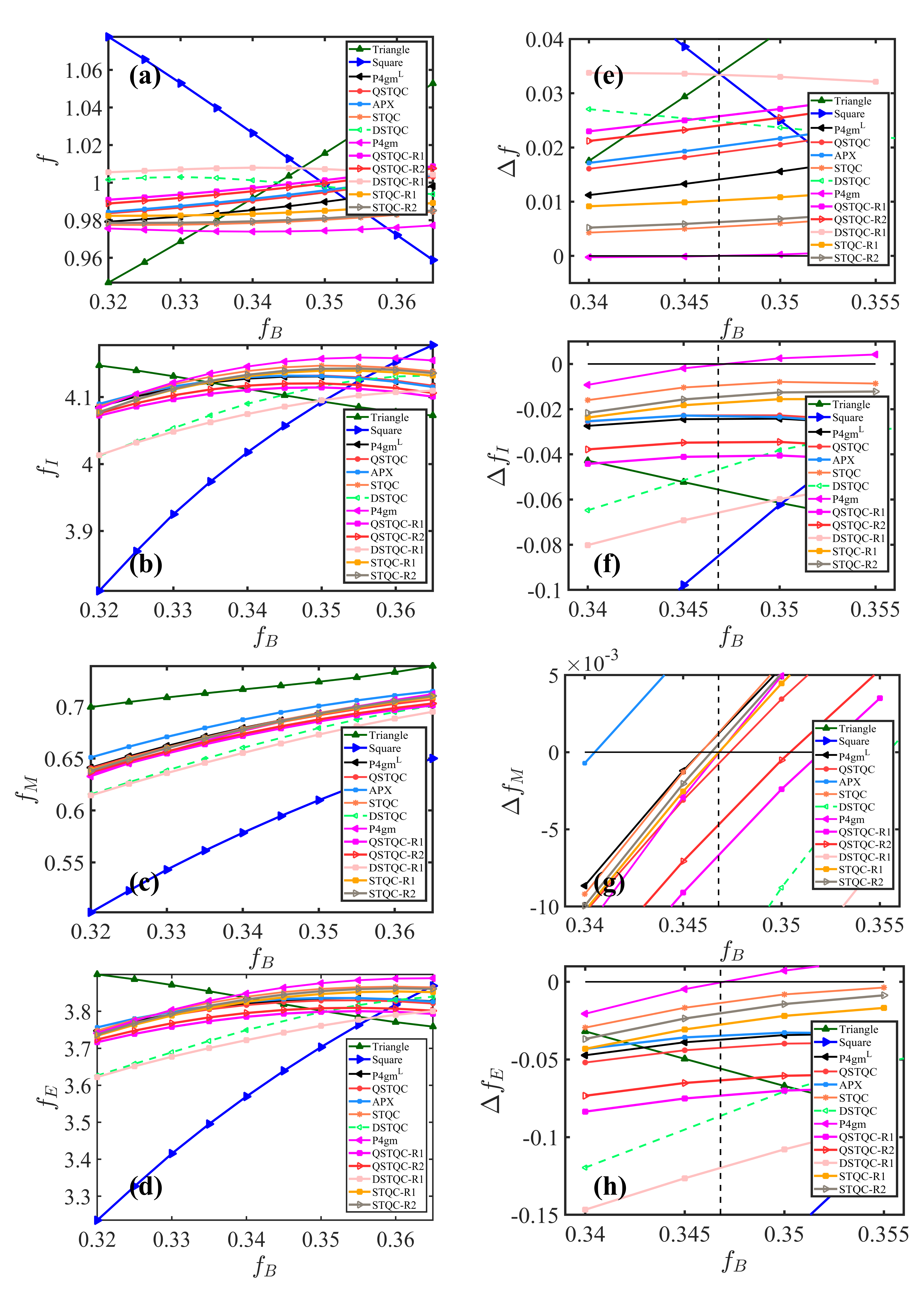}
	\caption{(a)-(d) The free energy comparison among all phases with $\eta=0.30, \chi
	=\chi_{AR}=0.30$. (e) Corresponding energy difference $\Delta f
=f-f_{P4gm}$ of (a), where $f_{P4gm}$ is the free energy
of P4gm on Triangle/Square boundary indicated by dashed lines. (f)-(h) Three
parts energy differences corresponds (e). }
	\label{fig:eeta}
\end{figure}

Three contributions of all phases when $\chi = 0.30, \eta=0.30$ together with the free energy difference reveal that entropic contribution $f_E$ is the dominant factor in thermodynamic stability, as shown in Fig.\,\ref{fig:eeta}.

\section{Diffraction peaks}
\label{sec:ddqcs}
Here, we supplement detailed information about the diffraction peaks.
In theoretical calculations, the Fourier coefficient at each diffraction peak is known. For any complex number $z = |z|e^{i\varphi}$, where the phase angle $\varphi \in [0,2\pi)$ can be used to characterize different diffraction peaks. The phase angles of monomer $A$ in the first-order diffraction peaks are shown in Tab.\,\ref{tab:phases}, at least the phase angles of three first-order diffraction peaks $(m=5, 8, 11)$ can distinguish between DDQCs and their approximants. 
\begin{table}[!htbp]
	\centering
	\caption{The phase angle $\varphi$ for DDQCs and their approximants when $\chi=0.30, f_B=0.35$.}
	\begin{tabular}{ccccccc}
	\toprule
	$m$&STQC &DSTQC
	&QSTQC &$\rm{P4gm^L}$ & P4gm & APX\\
	     \midrule
1&172.69 & 171.33& 167.43 &348.85&100.91 &351.20\\
2&351.54 & 349.84& 342.38 &166.59&241.35 &349.48\\
3&172.68 & 171.38& 169.83 &168.81& 61.34 &350.55\\
4&175.75 & 175.14& 169.15 &353.26&100.89 &355.48\\
5&359.97 &   0.15&   2.20 &179.95&320.43 &359.35\\
6&184.21 & 185.03& 186.96 &186.66&219.53 &  6.00\\
7&187.30 & 188.66& 192.56 & 11.14&259.08 &  8.79\\
8&  8.45 &  10.15&  17.61 &193.40&118.64 & 10.51\\
9&187.31 & 188.61& 190.16 &191.18&298.65 &  9.44\\
10&184.24& 184.85& 190.84 &  6.73&259.10 &  4.51\\
11&0.02  & 359.84& 357.79 &180.04& 39.56 &  0.64\\
12&175.78& 174.96& 173.03 &173.33&140.46 &353.99\\
\bottomrule
\end{tabular}
\label{tab:phases}
\end{table}

The angle deviation $\theta=|\theta_P-\theta_{STQC}|$ and length deviation
$d=r_{P}/r_{STQC}$ serve as metrics for quantifying the positional discrepancies of diffraction peaks, thereby providing criteria for distinguishing between the three DDQCs and their approximants. $(\theta_P,r_P)$ is the corresponding polar coordinate of $P=\left\{\rm{DSTQC, QSTQC}\right.$, $\left.  \rm{P4gm^L}, \rm{P4gm, APX}\right\}$ structures. Tab.\,\ref{tab:peaksA} clearly shows the first- and second-order diffraction peak deviations of block $A$. This demonstrates that the position of the first-order diffraction peaks alone is not sufficient to distinguish between the three DDQCs and their approximants. However, when combined with the second-order diffraction peaks, it becomes effective.

\begin{table}[H]
	\centering
    \caption{The diffraction peak deviation of block $A$ when $\chi=0.30, f_B=0.35$.}
	\begin{tabular}{ccccccccccc}
\toprule
	\multirow{2}*{$m$}&\multicolumn{2}{c}{DSTQC} &\multicolumn{2}{c}{QSTQC}
	&\multicolumn{2}{c}{$\rm{P4gm^L}$} &\multicolumn{2}{c}{P4gm} & \multicolumn{2}{c}{APX}\\
    \cmidrule(lr){2-11}
	& $\theta (^{\circ})$ & $d (\%)$ & $\theta (^{\circ})$ & $d (\%)$ & $\theta (^{\circ})$ & $d (\%)$  & $\theta (^{\circ})$ & $d (\%)$ & $\theta (^{\circ})$ & $d (\%)$ \\
    \midrule
	1 & 0	  & 0	  & 0	  & 1.78  & 0.89  &  2.01 & 0.07  & 2.38  & 2.21  & 2.85 \\
	2 & 0	  & 0	  & 0	  & 1.78  & 0     &  4.75 & 3.43  & 8.91  & 8.54  & 6.09 \\
	3 & 0	  & 0	  & 0	  & 1.78  & 0.89  &  2.01 & 3.36  & 9.14  & 1.53  & 11.45 \\
	4 & 0	  & 0	  & 0	  & 1.78  & 0.89  &  2.01 & 0.07  & 2.38  & 1.53  & 11.45 \\
	5 & 0	  & 0	  & 0	  & 1.78  & 0     &  4.75 & 3.43  & 8.91  & 8.54  & 6.09 \\
	6 & 0	  & 0	  & 0	  & 1.78  & 0.89  &  2.01 & 3.36  & 9.14  & 2.21  & 2.85 \\
	7 & 0	  & 0	  & 0	  & 1.78  & 0.89  &  2.01 & 0.07  & 2.38  & 2.21  & 2.85 \\
	8 & 0	  & 0	  & 0	  & 1.78  & 0     &  4.75 & 3.43  & 8.91  & 8.54  & 6.09 \\
	9 & 0	  & 0	  & 0	  & 1.78  & 0.89  &  2.01 & 3.36  & 9.14  & 1.53  & 11.45 \\
  10 & 0	  & 0	  & 0	  & 1.78  & 0.89  &  2.01 & 0.07  & 2.38  & 1.53  & 11.45\\
  11 & 0	  & 0	  & 0	  & 1.78  & 0     &  4.75 & 3.43  & 8.91  & 8.54  & 6.09\\
  12 & 0	  & 0	  & 0	  & 1.78  & 0.89  &  2.01 & 3.36  & 9.14  & 2.21  & 2.85\\
  13 & 1.67  & 5.14  & 12.81 & 31.73 & 9.63  & 28.55 & 20.61 & 60.71 & 6.01  & 4.48\\
  14 & 1.35  & 8.98  & 13.39 & 24.55 & 13.17 & 20.43 & 1.60  & 29.79 & 24.30 &22.42\\
  15 & 30.25 & 18.27 & 15.25 & 31.73 & 0     & 9.62  & 18.69 & 60.71 & 0     &21.91\\
  16 & 28.64 & 5.63  & 16.35 & 23.97 & 13.17 & 20.43 & 17.08 & 78.98 & 31.60 &30.44\\
  17 & 28.07 & 5.88  & 16.97 & 32.26 & 9.63  & 28.55 & 1.92  & 36.96 & 6.01  &4.48\\
  18 & 0     & 6.66  & 15.25 & 22.63 & 0     & 2.42  & 18.69 & 85.13 & 34.09 &21.05\\
  19 & 1.67  & 5.14  & 12.81 & 31.73 & 9.63  & 28.55 & 1.92  & 60.71 & 6.01  &4.48\\
  20 & 1.35  & 8.98  & 13.39 & 24.55 & 13.17 & 20.43 & 1.60  & 29.79 & 24.30 &22.42\\
  21 & 30.25 & 18.27 & 15.25 & 31.73 & 0     & 9.62  & 18.69 & 60.71 & 0     &21.91\\
  22 & 28.64 &  5.63 & 16.35 & 23.97 & 13.17 & 20.43 & 17.08 & 78.98 & 31.60 &30.44\\
  23 & 28.07 &  5.88 & 16.97 & 32.26 & 9.63  & 28.55 & 1.97  & 36.96 & 6.01  &4.48\\
  24 & 0     &  6.66 & 15.25 & 22.63 & 0     & 2.42  & 18.69 & 82.13 & 34.09 &21.05\\
  \bottomrule
\end{tabular}
\label{tab:peaksA}
\end{table}
 
\bibliography{TQC}